\documentclass[prd,nofootinbib,floats,superscriptaddress,eqsecnum,tightenlines,11pt]{revtex4}

\usepackage{hyperref}
\usepackage{graphicx}
\usepackage{setspace}
\usepackage{color}
\usepackage{amsmath,amssymb,amsfonts,amsthm,latexsym,stmaryrd}

\def\be{\begin{equation}}
\def\ee{\end{equation}}
\def\bes{\begin{equation*}}
\def\ees{\end{equation*}}
\def\ba{\begin{eqnarray}}
\def\ea{\end{eqnarray}}

\def\eps{\varepsilon}

\def\Tr{\text{Tr}}
\def\tr{\text{tr}}

\def\de{\mathrm{d}}

\newcommand{\f}{\frac}
\newcommand{\lb}{\big\lbrace}
\newcommand{\rb}{\big\rbrace}
\newcommand{\SU}{\text{SU}}
\newcommand{\SO}{\text{SO}}
\newcommand{\SP}{\text{Spin}}
\newcommand{\su}{\mathfrak{su}}
\newcommand{\so}{\mathfrak{so}}
\newcommand{\p}{{}^{\text{\tiny{($+$)}}\!}}
\newcommand{\m}{{}^{\text{\tiny{($-$)}}\!}}
\newcommand{\PM}{{}^{\text{\tiny{($\pm$)}}\!}}

\newtheorem{proposition}{Proposition}

\begin{document}

\title{Testing the imposition of the Spin Foam Simplicity Constraints}

\author{Marc Geiller}
\email{mgeiller@apc.univ-paris7.fr}
\affiliation{Laboratoire APC -- Astroparticule et Cosmologie, Universit\'e Paris Diderot Paris 7, 75013 Paris, France}
\author{Karim Noui}
\email{karim.noui@lmpt.univ-tours.fr}
\affiliation{Laboratoire de Math\'ematiques et Physique Th\'eorique, Universit\'e Fran\c cois Rabelais, Parc de Grandmont, 37200 Tours, France}
\affiliation{Laboratoire APC -- Astroparticule et Cosmologie, Universit\'e Paris Diderot Paris 7, 75013 Paris, France}

\begin{abstract}
We introduce a three-dimensional Plebanski action for the gauge group SO(4). In this model, the $B$ field satisfies quadratic simplicity constraints similar to that of the four-dimensional Plebanski theory, but with the difference that the $B$ field is now a one-form. We exhibit a natural notion of ``simple one-form'', and identify a gravitational sector, a topological sector and a degenerate sector in the space of solutions to the simplicity constraints. Classically, in the gravitational sector, the action is shown to be equivalent to that of three-dimensional first order Riemannian gravity. This enables us to perform the complete spin foam quantization of the theory once the simplicity constraints are solved at the classical level, and to compare this result with the various models that have been proposed for the implementation of the constraints after quantization. In particular, we impose the simplicity constraints following the prescriptions of the so-called BC and EPRL models. We observe that the BC prescription cannot lead to the proper vertex amplitude. The EPRL prescription allows to recover the expected result when, in this three-dimensional model, it is supplemented with additional secondary second class constraints.
\end{abstract}

\maketitle

\section*{Introduction}

\noindent Spin foam models \cite{baez,oriti-thesis,perez-review,rovelli-book} constitute an exciting proposal for the definition of a background independent and non-perturbative quantization of general relativity. They were introduced originally \cite{reisenberger-rovelli2} as a way to implement the dynamics of loop quantum gravity \cite{thiemann-book,ashtekar-lewandowski}, and can be thought of as representing a sum over histories of the gravitational field \cite{reisenberger-rovelli}.

In three spacetime dimensions, pure gravity being a topological theory (in the sense that it has no local degrees of freedom), spin foam models can be obtained as an exact simplicial path integral for the first-order Palatini action, where the gauge group is taken to be $\SU(2)$ in Riemannian signature (the whole symmetry group is larger than $\SU(2)$ and depends on the sign of the cosmological constant $\Lambda$). This corresponds to the celebrated Ponzano-Regge \cite{ponzano-regge} and Turaev-Viro \cite{turaev-viro} models when $\Lambda=0$ and $\Lambda>0$ respectively. Their vertex amplitudes are simply given by $\SU(2)$ and $\text{U}_q(\su(2))$ (for $q$ a root of unity) $6j$ coefficients. It has been shown \cite{noui-perez} that the Ponzano-Regge amplitudes can be obtained from a canonical quantization of $2+1$ gravity in the spirit of loop quantum gravity, in which the kinematical states are given by spin networks, and the dynamics leads to a spin foam evolution. The Turaev-Viro amplitudes can also be obtained as the scalar product between physical states which are in turn given in terms of unitary representations of some quantum groups closely related to $\text{U}_q(\su(2))$ \cite{witten, reshetikhin-turaev}. Therefore, in three dimensions, there is a clear understanding of the relationship between the so-called canonical and covariant approaches to quantum gravity (at least when the signature is Riemannian).

In four spacetime dimensions, the construction of such a correspondence between the canonical and spin foam quantizations is still an open problem \cite{revue,dupuis-livine,rovelli-speziale}. Since it is technically too involved to follow the construction of the three-dimensional theory and perform the straight simplicial path integral quantization of the four-dimensional Palatini action (see \cite{BFT} for an attempt to follow this direction), spin foam models have to be derived using an alternative strategy. More precisely, the approach to four-dimensional spin foam models is based on the fact that gravity can be formulated as a constrained topological field theory, by virtue of the so-called Plebanski action \cite{plebanski}. This latter is defined as the sum of a well-known topological $BF$ action \cite{horowitz}, plus a set of constraints on the $B$ field ensuring on-shell the condition that it comes from the exterior product of two tetrad one-form fields\footnote{There is also a topological sector and a degenerate sector in the solutions to the simplicity constraints, but we shall come back to this later on.}. With this solution for the $B$ field, the Plebanski action reduces at the classical level to the usual Hilbert-Palatini action of first order general relativity. To derive a spin foam model, the following general strategy is adopted. First, the topological $BF$ theory is discretized exactly on a two-complex, and the basic variables, i.e. the $B$ field and holonomies of the spacetime connection, are promoted to quantum operators. Then, the simplicity constraints are imposed at the quantum level as restrictions on group-theoretical data. The main challenge in this approach is essentially to impose consistently these second class simplicity constraints, and several proposals defining the different available spin foam models have been put forward in order to do so.

The three spin foam models for four-dimensional quantum gravity that have been studied the most are the Barrett-Crane (BC) model \cite{BC}, the Engle-Pereira-Rovelli-Livine (EPRL) model \cite{EPR,EPRL,livine-speziale1}, and the Freidel-Krasnov (FK) model \cite{FK,livine-speziale2}. They all rely on the general strategy outlined above, even though their construction requires to impose the simplicity constraints in drastically different ways. In the BC model, which was the first four-dimensional model to be introduced, the simplicity constraints are imposed as strong operator relations. This has the result of assigning simple representations of the gauge group to faces of the dual triangulation $\Delta^*$, and a specific unique intertwiner, known as the BC intertwiner \cite{BC,reisenberger-BC}, to edges of $\Delta^*$. Due to difficulties in reproducing the correct semiclassical limit \cite{scBC1,scBC2,scBC3} and the structure of the graviton propagator \cite{alesci-rovelli}, this model was partly discarded\footnote{See \cite{baratin-oriti,baratin-oriti2} for an overview of the criticisms and recent arguments in favor of a reconsideration of the model.}, and the search for new models was motivated with the additional hope to relate the spin foam quantization to the canonical structure of loop quantum gravity. This search culminated with the introduction of the EPRL and FK models, which are both based on the introduction of a linear version of the (originally quadratic) simplicity constraints \cite{FK}, and take as an important additional input the inclusion of the Barbero-Immirzi parameter. In these two models, the over-imposition of the simplicity constraints which is thought to be responsible for the issues with the BC vertex, is cured by using a weak imposition by means of coherent states for the FK model, or the Gupta-Bleuler scheme for the EPRL model. Our goal here is not to review extensively the details of these constructions, but rather to focus on the implementation of the simplicity constraints by introducing a model which allows for an immediate comparison between the alternative schemes. In fact, it has been argued on several occasions \cite{alexandrov1,alexandrov2,alexandrov3} that the current spin foam models miss the imposition of additional secondary second class constraints that are generated by the usual simplicity constraints. The usual point of view in the derivation of spin foam models is that it might be sufficient to impose the primary simplicity constraints consistently at all times, since the secondary constraints arise in fact as a consequence of this requirement. But even if this is true at the classical level, at the quantum level the secondary constraints are allowed to have non-vanishing fluctuations, and ought therefore be imposed as well. It is however very hard in the four-dimensional case to implement concretely this idea. In other words, we understand that there is a difficulty which needs to be resolved in the current four-dimensional spin foam models, but even if we have generic arguments indicating how to do so, a concrete realization has never been put forward.

Let us summarize the situation. At the classical level, it is clear that the Plebanski action is equivalent to the Palatini action once the $B$ field is forced to be simple. But since we do not know how to compute the simplicial path integral for the Palatini action, we try to compute that of the Plebanski action by first quantizing the unconstrained topological theory and then imposing the simplicity constraints at the quantum level. In fact, it is quite clear that we are restricted in our understanding of how to properly perform this last step, by the fact that we do no know what to ``expect'' from the spin foam quantization of the four-dimensional Plebanski action. For this reason, it would be nice to have a model in which we could work out explicitly the two alternative methods. By this, we mean that we would like to perform the spin foam quantization of the action in which the simplicity constraints have already been solved at the classical level, and then compare this result to the spin foam quantization in which the simplicity constraints are solved in the quantum theory. This would be a way to test the various proposals to deal with the simplicity constraints. It has already been argued in the literature \cite{alexandrov-roche} that dealing with second class constraints at the quantum level can lead to inconsistencies, and is not compatible with the quantization program \`a la Dirac. However, the argument alone is not fully convincing, since it was formulated on a finite dimensional model with no clear analogy with gravity, and since it does not give a clear explanation for how the various impositions of the constraints in the four-dimensional spin foam models should lead to a model with inconsistent physical predictions. It might very well be that despite the fact that we are dealing with second class constraints at the quantum level, there exists a preferred scheme in which the simplicity constraints can be imposed in a (yet-to-be defined) robust way to lead to an acceptable model for quantum gravity. In the end, the only way to discriminate between the various proposals is either to extract physical predictions to compare with experiments, or at least to test the strategies on toy models which bear a close analogies with gravity. Let us also point out that a model was introduced in \cite{alexandrov2} in order to illustrate the claim that additional secondary second class constraints should be taken into account in the spin foam models. This model is based on the idea of reducing a four-dimensional $\SO(4)$ $BF$ theory to an $\SU(2)$ one by means of simplicity constraints, and demonstrates that this is only possible if certain secondary second class constraints are imposed. The weak point of this model is that the constraints that it imposes are not derived from the Hamiltonian analysis of a given Plebanski theory, but are instead put in by hand.

\begin{figure}[h]
%\begin{diagram}
%S_\text{BF}&\rTo^{\mathcal{C}=0}&S_\text{grav}\\
%\dTo&&\dTo\\
%\mathcal{Z}_\text{BF}&\rTo^{~~\text{BC or EPRL ?}~~}&\mathcal{Z}_\text{grav}
%\end{diagram}
\includegraphics{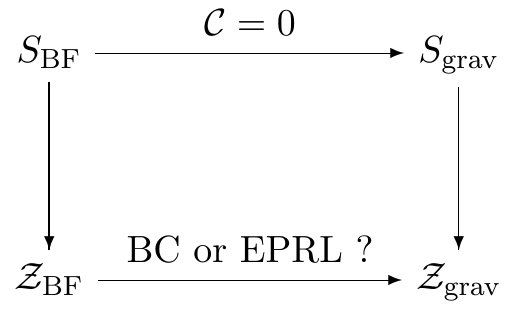}
\caption{The starting point of this investigation is the three-dimensional action $S_\text{BF}$. After imposition of the simplicity constraints $\mathcal{C}$, it reduces to the usual Hilbert-Palatini action, whose discretized partition function $\mathcal{Z}_\text{grav}$ can be computed explicitly. The question we want to ask is whether it is possible to derive the same vertex amplitude by imposing the simplicity constraints at the quantum level on the partition function $\mathcal{Z}_\text{BF}$ for the unconstrained action. In other words, does the imposition of the simplicity constraints \`a la BC or EPRL render this diagram commutative? If not, what is the additional input which is necessary in order to reconstruct the Ponzano-Regge model from the spin foam quantization of the three-dimensional Plebanski action?}
\label{diagram}
\end{figure}

The aim of this paper is to formulate a robust model which allows to test the imposition of the spin foam simplicity constraints. Since pure gravity is always topological in three spacetime dimensions, it makes a priori no sense to have a Plebanski formulation in which the $B$ field is forced to be simple. In fact three-dimensional $BF$ theory is already equivalent to gravity, when $B$ is a one-form field valued in the Lie algebra of $\SU(2)$. However, if one replaces the gauge group $\SU(2)$ by $\SO(4)$ (or its double cover $\SP(4)$), the three-dimensional theory admits simplicity constraints $\mathcal{C}$ similar to that of the four-dimensional theory. In the gravitational sector of solution to these simplicity constraint, the action becomes that of first order Riemannian gravity. In the topological sector, the theory is trivial. Therefore, at the classical level, the Plebanski theory that we introduce reduces to the Hilbert-Palatini action, whose spin foam quantization naturally leads to the Ponzano-Regge amplitudes. Now, it is also possible to perform the spin foam quantization of the three-dimensional $\SO(4)$ $BF$ theory, and to impose the simplicity constraints at the quantum level, mimicking in particular the prescription of the BC and EPRL\footnote{Strictly speaking, we are going to study the EPR model \cite{EPR} because we do not consider for the moment the inclusion of the Barbero-Immirzi parameter.} models. This allows for a direct verification of whether any of these proposals leads to the proper vertex amplitude. Our construction can be summarized in the diagram of figure \ref{diagram}.

We start in section \ref{sec1} by introducing the classical framework for our analysis. In particular, we show how the three-dimensional $\SO(4)$ $BF$ action is reduced by the simplicity constraints to the usual action for three-dimensional gravity, and we perform the Hamiltonian analysis of the Plebanski theory to support this fact. Section \ref{sec2} is devoted to the study of the quantum theory. We first recall the state sum model corresponding to the constrained classical action, and then perform the spin form quantization of the Plebanski theory. For this, we start by writing down the simplicial path integral for the $\SO(4)$ $BF$ theory, and then study the imposition of the simplicity constraints. In particular, we study the BC and EPRL prescriptions, in which the constraints are imposed strongly and weakly (in a sense that we will make more precise), respectively. We also study the so-called Warsaw modification \cite{KKL1,KKL2,BHKKL} of the EPRL model in order to illustrate its signification on the three-dimensional model at hand. Our conclusion is that the BC prescription cannot lead to the proper vertex amplitude, whereas the EPRL prescription, when supplemented with the imposition of the additional secondary second class constraints, enables one to recover the expected result. This might be a strong indication of the fact that a weak imposition of the simplicity constraints in the quantum theory is an important ingredient in order to derive the vertex amplitude of four-dimensional quantum gravity. Nevertheless, we want to emphasize that in our three-dimensional model, none of above-mentioned proposals lead by themselves to the proper vertex amplitude, which is that of the Ponzano-Regge model. The reason for this is clear: the current spin foam models miss the imposition of the secondary second class constraints that are exhibited in the canonical analysis of the Plebanski theory. Indeed, we show how these constraints can be solved by imposing conditions on the spatial components of the connection, and how their imposition in the path integral with the EPRL prescription allows to recover the proper vertex amplitude. 

\section{Classical theory -- actions for three-dimensional gravity}
\label{sec1}

\noindent To introduce our model, let us start with the familiar $BF$ action, written in three dimensions and for the gauge group $\SO(4)$. In terms of an $\so(4)$-valued one-form field $B^{IJ}_\mu$ (our notations are defined in appendix \ref{notations}) and the spacetime connection $\omega^{IJ}_\mu$, it is given by
\be\label{BF action}
S_\text{BF}[B,\omega]
=\f{1}{2}\int_\mathcal{M}\de^3x\,\eps^{\mu\nu\rho}\,\Tr(B_\mu,F_{\nu\rho})
=\f{1}{2}\int_\mathcal{M}\de^3x\,\eps^{\mu\nu\rho}B_{\mu IJ}F_{\nu\rho}^{IJ},
\ee
where $\Tr(\cdot\,,\cdot)$ denotes a trace in the Lie algebra $\so(4)$, and $F_{\mu\nu}$ is the curvature two-form associated to the connection. Clearly, this action is not that of Riemannian three-dimensional first order gravity, but this latter (or at least two copies of it) can be obtained if we use a decomposition of the fields into self-dual and anti self-dual components. Here we are not interested in doing this decomposition, because our goal is to mimic the constrained four-dimensional action which is the starting point for the spin foam quantization. To see that this is indeed possible, let us proceed in two steps. First, let us assume that the $B$ field is of the form
\be\nonumber
B^{IJ}_\mu=\eps^{IJ}_{~~KL}\chi^Ke^L_\mu,
\ee
where $\chi^K$ is a zero-form in $\mathbb{R}^4$, and $e^L_\mu$ a Lie algebra-valued one-form. This is the three-dimensional analogue of the notion of simple two-form field that is used in four-dimensional Plebanski theory. Second, let us choose a gauge (which we will refer to as the time-gauge) in which $\chi^I=(\chi^0,0,0,0)$. This gauge choice reduces the symmetry group $\SO(4)$ to its $\SU(2)$ subgroup. A key observation is that it is always possible to fix $\chi^0=1$ due to the presence of the ``scaling'' symmetry
\be\label{rescaling}
\chi^I\longrightarrow\alpha\chi^I,\qquad e^I_\mu\longrightarrow\f{1}{\alpha}e^I_\mu,
\ee
for $\alpha\neq0$. With this choice of gauge, the action (\ref{BF action}) becomes the usual action for gravity\footnote{This is in fact not straightforward, since the curvature $F^{ij}_{\mu\nu}$ in (\ref{gravity action}) involves connection components of the type $\omega^{0i}_\mu$. However, one can see that these terms appear quadratically as Lagrange multipliers, and that they are therefore vanishing by virtue of the equations of motion.}
\be\label{gravity action}
S_\text{BF}[B(e),\omega]=S_\text{grav}[e,\omega]=\f{1}{2}\int_\mathcal{M}\de^3x\,\eps^{\mu\nu\rho}\eps_{ijk}e^i_\mu F^{jk}_{\mu\nu}.
\ee
We have included the first equality to make explicit the fact that the three-dimensional Hilbert-Palatini action is simply a constrained $BF$ action. This argument suggests that it is possible to obtain the action for gravity starting from (\ref{BF action}) if we impose a simplicity-like condition on the $B$ field, and pick up a rotational $\SU(2)$ subgroup of $\SO(4)$ by making a gauge choice for $\chi^I$ (notice that this choice is not canonical).

We are going to see in this section that the simplicity of the $B$ field can be enforced with a Plebanski action whose Hamiltonian analysis will be detailed.

\subsection{Plebanski action and simplicity constraints}

\noindent As we mentioned above, the first things that we would like to do is to impose the simplicity of the $B$ field in the $\SO(4)$ $BF$ theory. Following what is done in four dimensions, we are going to introduce the Plebanski action as the sum of the $BF$ action (\ref{BF action}) plus a set of constraints on the $B$ field. Our aim is to reproduce simplicity-like conditions ensuring that we are indeed dealing with three-dimensional gravity. The action that we consider is\footnote{Notice that the constraint term is not a three-form since the spacetime indices of the $B$ field are contracted with those of the multiplier $\phi$, and not those of the antisymmetric Levi-Civita symbol. This is similar to what happens in the four-dimensional Plebanski theory when the multiplier acts on the spacetime indices of the $B$ field.}
\be\label{3d plebanski}
S_\text{Pl}[B,\omega,\phi]
=S_\text{BF}[B,\omega]+~\text{constraints}
=\f{1}{2}\int_\mathcal{M}\de^3x\,\Big(\eps^{\mu\nu\rho}\,\Tr(B_\mu,F_{\nu\rho})+\phi^{\mu\nu}\,\Tr(B_\mu,\star B_\nu)\Big),
\ee
where $\star$ denotes the Hodge dual operator in $\so(4)$, and $\phi^{\mu\nu}$ is a symmetric Lagrange multiplier used to enforce the constraints
\be\label{simplicity}
\mathcal{C}_{\mu\nu}\equiv\Tr(B_\mu,\star B_\nu)=\f{1}{2}\eps_{IJKL}B^{IJ}_\mu B^{KL}_\nu\approx0.
\ee
We want to understand the exact meaning of these constraints in the present context. The key difference with four-dimensional Plebanski theory is that we are dealing here with vectors, and not bivectors. Therefore, we need another notion of simplicity than the usual one used in spin foam models. As we have already mentioned earlier, we are going to say that a vector $B^{IJ}_\mu$ is simple if it can be written as
\be\nonumber
B^{IJ}_\mu=\eps^{IJ}_{~~KL}\chi^Ke^L_\mu,
\ee
where $\chi^K$ is a zero-form in $\mathbb{R}^4$, and $e^L_\mu$ a one-form. The fact that the constraint $\mathcal{C}$ defined by (\ref{simplicity}) can be used to obtain simple vectors is ensured by the following proposition:
\begin{proposition}\label{prop}
When the simplicity constraints $\mathcal{C}$ are satisfied, there are three possible solutions for the $B$ field. They correspond respectively to:
\begin{itemize}
\item A gravitational sector, in which there exists an $\mathbb{R}^4$-valued one form $e^I_\mu$ and a vector $\chi^I$ in $\mathbb R^4$ such that
$$B^{IJ}_\mu=\eps^{IJ}_{~~KL}\chi^Ke^L_\mu,
\qquad
K^i_\mu\equiv B^{0i}_\mu=(\chi\wedge e_\mu)^i,
\qquad
L^i_\mu\equiv\f{1}{2}\eps^i_{~jk}B^{jk}_\mu=\chi^0e^i_\mu-\chi^ie^0_\mu.$$
\item A topological sector, in which there exists an $\mathbb R^4$-valued one form $e^I_\mu$ and a vector $\chi^I$ in $\mathbb R^4$ such that
$$B^{IJ}_\mu=\star\eps^{IJ}_{~~KL}\chi^Ke^L_\mu=\chi^Ie^J_\mu-\chi^Je^I_\mu,
\qquad
K^i_\mu=\chi^0e^i_\mu-\chi^ie^0_\mu,
\qquad
L^i_\mu=(\chi\wedge e_\mu)^i.$$
\item A degenerate sector, in which $\det(K_\mu^i)=0$ and $\det(L_\mu^i)=0$.
\end{itemize}
\end{proposition}
\begin{proof}
Let us introduce the boost and rotational components of the field $B^{IJ}_\mu$. They are given respectively by $K^i_\mu$ and $L^i_\mu$. Each of these one-forms can be interpreted as a vector in $\mathbb R^3$, and the simplicity constraints simply mean that the vectors $K_\mu$ are orthogonal to the vectors $L_\nu$, i.e.
\be\label{simplicitygeom}
L_\mu\cdot K_\nu=0,\qquad\forall\,\mu,\nu.
\ee
Let us first assume that the three vectors $L_\mu$ form a basis of $\mathbb R^3$. Due to the simplicity constraints (\ref{simplicitygeom}), the three remaining vectors $K_\mu$ are linked, and therefore they lie in the same plane, whose (non-necessarily unit) normal is denoted by $\chi$. More precisely, there exist three vectors $e_\mu$ such that
\be\label{Kgrav}
K_\mu=\chi\wedge e_\mu.
\ee
Furthermore, since the vectors $L_\mu$ and $K_\mu$ are orthogonal, we can write
to
\be\label{Bgrav}
L_\mu=\chi^0e_\mu-e^0_\mu\chi,
\ee
where $\chi^0$ is a (non-vanishing) scalar, and $e^0_\mu$ is a real-valued one form. Due to the fact that the vectors $L_\mu$ form a basis, the three vectors $e_\mu$ form a basis as well. The solution given by (\ref{Kgrav}) and (\ref{Bgrav}) corresponds to the gravitational sector.

If we assume on the contrary that the three vectors $K_\mu$ form a basis, the same construction can be used to obtain the topological sector described in proposition \ref{prop}.

When $\det(K_\mu^i)=0$ and $\det(L_\mu^i)=0$, i.e. neither the $K_\mu$ nor the $L_\mu$ form a basis, we say that the $B$ field belongs to the degenerate sector.
\end{proof}

It is clear from the expression of $B^{IJ}_\mu$ in the gravitational sector that in the time gauge, where we have $\chi^i=0$ and $\chi^0=1$, the boost component $K^i_\mu$ is vanishing, while the rotational component $L^i_\mu$ reduces to the triad $e^i_\mu$. This supports the heuristic argument that we have given at the beginning of this section in order to derive the action (\ref{gravity action}) from the $\SO(4)$ $BF$ theory. Note also that in the gravitational sector we have $\det(K_\mu^i)=0$, while in the topological sector we have $\det(L_\mu^i)=0$.

Let us conclude with a remark. There are 6 simplicity constraints $\mathcal{C}_{\mu\nu}$ acting on the 18 components $B^{IJ}_\mu$ of the $B$ field. Therefore we expect the simple $B$ field to be written in terms of 12 components only. Here, we have expressed $B$ in terms of the 16 components $\chi^I$ and $e^I_\mu$. However, notice that there are 4 redundant components due to the presence of the following symmetries:
\begin{itemize}
\item A rescaling symmetry, given by (\ref{rescaling}), which allows to remove one component (for instance, we can fix $\chi^I\chi_I=1$).
\item Three translational symmetries acting in the non-degenerate sector. They are generated by a vector $\beta_\mu$, and act like
\be\nonumber
e^i_\mu\longrightarrow e^i_\mu+\beta_\mu\chi^i,\qquad e^0_\mu\longrightarrow e^0_\mu+\beta_\mu\chi^0.
\ee
As a consequence, we can always choose $e_\mu^0=0$, but here we do not make this choice in order not to break the $\SO(4)$ symmetry of the theory.
\end{itemize}

\subsection{Gravitational and topological sectors}

\noindent Now that we have a good understanding of how the simplicity constraints can be used to define a sector of the Plebanski theory in which the vector $B^{IJ}_\mu$ is simple, let us see how it is possible to recover the action for three-dimensional gravity. For this, we are going to study separately the gravitational and topological sectors introduced in proposition (\ref{prop}).

Using the decomposition of $\so(4)$ into self-dual and anti self-dual generators (see appendix \ref{so4}), we can write the $B$ field as
\be\nonumber
B^{IJ}_\mu=\p B^i_\mu\p J^{IJ}_i+\m B^i_\mu\m J^{IJ}_i,
\ee
where $\PM B^i_\mu=\Tr(B_\mu,\PM J^i)$, and the $BF$ action (\ref{BF action}) therefore becomes
\be\label{BF selfdual}
S_\text{BF}[\p B,\m B,\p\omega,\m\omega]=\f{1}{2}\int_\mathcal{M}\de^3x\,\eps^{\mu\nu\rho}\Big(\Tr(\p B_\mu,\p F_{\nu\rho})+\Tr(\m B_\mu,\m F_{\nu\rho})\Big).
\ee
As usual, the equations of motion with respect to the connection $\PM\omega$ lead to the torsion-free condition
\be\nonumber
T(\PM B,\PM\omega)=0,
\ee
and if $\det(\PM B)\neq0$, this relation can be inverted to find the torsion-free spin connection. This latter, when plugged back into the original action (\ref{BF selfdual}), leads to the sum of two second order Einstein-Hilbert actions,
\be\label{einstein-hilbert}
S_\text{EH}[\p g_{\mu\nu},\m g_{\mu\nu}]=\f{1}{2}\epsilon^+\int_\mathcal{M}\de^3x\sqrt{|\p g|}\,\mathcal{R}(\p g_{\mu\nu})+\f{1}{2}\epsilon^-\int_\mathcal{M}\de^3x\sqrt{|\m g|}\,\mathcal{R}(\m g_{\mu\nu}),
\ee
each being defined with respect to a two-dimensional Urbantke-like metric \cite{urbantke} $\PM g_{\mu\nu}=\PM B_\mu\cdot\PM B_\nu$ (in the sense that it is constructed with the $B$ field). In this expression, $\epsilon^\pm$ denotes the sign of $\det(\PM B)$ (we refer the reader to appendix \ref{appdet} for the calculation of $\epsilon^\pm$).

In the gravitational and topological sectors, the self-dual and anti self-dual components of the $B$ field are given by:
%\begin{itemize}
%\item gravitational sector: $\quad\quad\PM B^i_\mu=\mp(\chi\wedge e_\mu)^i+(\chi^ie^0_\mu-\chi^0e^i_\mu),\qquad\epsilon^+=\epsilon^-$,
%\vspace{-0.2cm}
%\item topological sector: $~~\,\quad\quad\PM B^i_\mu=-(\chi\wedge e_\mu)^i\pm(\chi^ie^0_\mu-\chi^0e^i_\mu),\qquad\epsilon^+=-\epsilon^-$.
%\end{itemize}
%\vspace{-1.85cm}
%\begin{subequations}
%\ba
%~\label{selfdual BG}\\
%~\label{selfdual BT}
%\ea
%\end{subequations}
\begin{subequations}
\ba
~\label{selfdual BG}\\
~\label{selfdual BT}
\ea
\end{subequations}
\vspace{-2.2cm}
\begin{itemize}
\item gravitational sector: $\quad\quad\PM B^i_\mu=\mp(\chi\wedge e_\mu)^i+(\chi^ie^0_\mu-\chi^0e^i_\mu),\qquad\epsilon^+=\epsilon^-$,
\vspace{-0.2cm}
\item topological sector: $~~\,\quad\quad\PM B^i_\mu=-(\chi\wedge e_\mu)^i\pm(\chi^ie^0_\mu-\chi^0e^i_\mu),\qquad\epsilon^+=-\epsilon^-$.
\end{itemize}
In each of these two sectors, we can now compute the two Urbantke metrics $\PM g_{\mu\nu}$ defined by the expressions above. In fact, from the simplicity constraints, we know already that $\p g_{\mu\nu}=\m g_{\mu\nu}$. It is however less trivial to see that the metrics in the topological and gravitational sector are identical. A simple calculation shows that we have
\be\label{urbantke metrics}
\p g_{\mu\nu}=\m g_{\mu\nu}\equiv g_{\mu\nu}=(e_\mu\cdot e_\nu)(\chi^2+(\chi^0)^2)-(\chi\cdot e_\mu)(\chi\cdot e_\nu)+\chi^2e^0_\mu e^0_\nu-\chi^0\chi\cdot(e^0_\mu e_\nu+e^0_\nu e_\mu),
\ee
where $\chi^2\equiv\chi^i\chi_i$. In the time gauge, this metric reduces to $g_{\mu\nu}=e_\mu\cdot e_\nu$.

Gathering these results on the Urbantke metric and the sign factors $\epsilon^\pm$, we can conclude that the action (\ref{BF action}) reduces in the gravitational sector to
\be\nonumber
S_\text{EH}[g_{\mu\nu}]=\int_\mathcal{M}\de^3x\sqrt{|g|}\,\mathcal{R}(g_{\mu\nu}),
\ee
while in the topological sector it becomes simply $S_\text{EH}[g_{\mu\nu}]=0$.

This shows that the gravitational sector to the solutions of the simplicity constraints $\mathcal{C}$ corresponds indeed to three-dimensional gravity, whereas in the topological sector the action vanishes on-shell. This is exactly what happens in the four-dimensional Plebanski theory. Interestingly, the presence of a topological sector in this three-dimensional theory allows for the introduction of a Barbero-Immirzi parameter, but we shall come back to this point in section \ref{sec3}.

\subsection{Canonical analysis of the Plebanski action}
\label{sec:canonical}

\noindent In this subsection we perform the canonical analysis of the three-dimensional $\SO(4)$ Plebanski theory. It is similar to the study of the four-dimensional Plebanski action \cite{BHNR}. Using a $2+1$ decomposition of the spacetime manifold, (\ref{3d plebanski}) becomes
\be\nonumber
S_\text{Pl}[B,\omega,\phi]
=\int_\mathbb{R}\de t\int_\Sigma\de^2x\,\bigg(-\eps^{ab}\,\Tr(B_a,\partial_0\omega_b)
+\f{1}{2}\eps^{ab}\,\Tr(B_0,F_{ab})
+\eps^{ab}\,\Tr(\omega_0,T_{ab})
+\f{1}{2}\phi^{\mu\nu}\,\Tr(B_\mu,\star B_\nu)\bigg),
\ee
where the curvature and torsion two-forms are defined by\footnote{The torsion is here defined up to a factor $1/2$ for convenience. Its exact expression is $\widetilde{T}_{ab}=\partial_aB_b-\partial_bB_a+[\omega_a,B_b]-[\omega_b,B_a]$.}
\be\label{curvature}
\eps^{ab}F_{ab}=\eps^{ab}(\partial_a\omega_b-\partial_b\omega_a+[\omega_a,\omega_b]),\qquad
\eps^{ab}T_{ab}=\eps^{ab}(\partial_aB_b+[\omega_a,B_b]).
\ee
We can see from this formula for the action that $\omega_0$, $B_0$ and $\phi^{\mu\nu}$ are non-dynamical variables, since the Lagrangian does not feature their time derivatives. The time component $\omega_0$ of the connection and $\phi^{\mu\nu}$ are true Lagrange multipliers, enforcing respectively the torsion-free and simplicity constraints. However, since $B_0$ is involved quadratically in the simplicity constraint, it cannot be treated as a Lagrange multiplier. For this reason, we add to the Lagrangian the term
\be\nonumber
\Tr(\pi_0,\partial_0B_0)+\Tr(\mu_0,\pi_0),
\ee
where $\pi_0$ and $\mu_0$ are new auxiliary $\SO(4)$-valued fields that do not affect the dynamics of the theory.

The basic variables of the theory are then the 12 spatial components $\omega_a^{IJ}$ of the connection, their 12 canonical momenta $B^{IJ}_a$, and the 6 components $B_0^{IJ}$ with their momenta $\pi_0^{IJ}$. The Poisson structure is given by
\be\label{phase space}
\lb B_a^{IJ}(x),\omega_b^{KL}(y)\rb=-\delta^{IJ,KL}\eps_{ab}\delta^2(x-y),\qquad\lb B_0^{IJ}(x),\pi_0^{KL}(y)\rb=\delta^{IJ,KL}\delta^2(x-y),
\ee
where $\delta^{IJ,KL}=(\delta^{IK}\delta^{JL}-\delta^{IL}\delta^{JK})/2$. The total Hamiltonian is
\be\label{hamiltonian}
H=-\int_\Sigma\de^2x\,\bigg(\f{1}{2}\eps^{ab}\,\Tr(B_0,F_{ab})
+\eps^{ab}\,\Tr(\omega_0,T_{ab})
+\f{1}{2}\phi^{\mu\nu}\,\Tr(B_\mu,\star B_\nu)
+\Tr(\mu_0,\pi_0)\bigg).
\ee
We can now identify the primary constraints and compute their Poisson bracket with this Hamiltonian in order to study their evolution in time.

\subsubsection{\textbf{Primary and secondary constraints}}

\noindent We have the following 18 primary constraints:
\be\nonumber
\mathcal{C}_{\mu\nu}=\Tr(B_\mu,\star B_\nu)\approx0,\qquad T_{ab}\approx0,\qquad\pi_0\approx0.
\ee
The first set is obtained by varying the action with respect to $\phi^{\mu\nu}$, the second set by varying the action with respect to $\omega_0$, and the third one by varying with respect to $\mu_0$

Before going any further, let us introduce smeared variables in order to deal with the delta functions appearing in the Poisson brackets. To this end, we define the quantities
\be\nonumber
T(u)=\int_\Sigma\de^2x\,\eps^{ab}\,\Tr(u,T_{ab}),\qquad F(u)=\int_\Sigma\de^2x\,\eps^{ab}\,\Tr(u,F_{ab}),
\ee
for any $\SO(4)$-valued smooth test function $u(x)$ on $\Sigma$. This enables us to compute
\be\nonumber
\lb F(u),F(v)\rb=0,\qquad\lb T(u),T(v)\rb=-T([u,v]),\qquad\lb T(u),F(v)\rb=-F([u,v]),
\ee
and
\be\nonumber
\lb T(u),B_a\rb=[u,B_a],\qquad\lb T(u),\omega_a\rb=-\mathcal{D}_au,\qquad\lb F(u),B_a\rb=-2\mathcal{D}_au,\qquad\lb F(u),\omega_a\rb=0,
\ee
where $\mathcal{D}$ stands for the covariant derivative with respect to the connection $\omega$. Now we can compute the Poisson bracket of the primary constraints with the total Hamiltonian, to see if the requirement that they be preserved under the time evolution gives rise to secondary constraints.

\subsubsection*{{a. Evolution of the constraints $T_{ab}$}}

\noindent First, it is easy to show that the conservation of $T_{ab}$ does not lead to any secondary constraints. To see this, we can introduce the new constraint
\be\label{SO(4) generator}
G(u)\equiv T(u)-\int_\Sigma\de^2x\,\Tr(u,[B_0,\pi_0]),
\ee
and notice that it is in fact the generator of the infinitesimal $\SO(4)$ gauge transformations on the phase space. Indeed, on the action of $G(u)$ on the phase space variables (\ref{phase space}) is given by
\be\nonumber
\lb G(u),B_\mu\rb=[u,B_\mu],\qquad\lb G(u),\omega_a\rb=-\mathcal{D}_au.
\ee
Since the Hamiltonian contains only terms with traces, it is left invariant under the action of this constraint, which therefore satisfies $\dot{G}=\lb H,G\rb\approx0$. The modification (\ref{SO(4) generator}) of the constraint is permissible since it amounts to adding a term which is itself proportional to a constraint.

\subsubsection*{{b. Evolution of the constraints $\pi_0$}}

\noindent The requirement that the primary constraint $\pi_0$ be preserved in time leads to the following relation:
\be\nonumber
\dot{\pi}_0=\lb H,\pi_0\rb=-\f{1}{2}\eps^{ab}F_{ab}-\phi^{0a}\star\!B_a-\phi^{00}\star\!B_0\approx0.
\ee
Among these 6 equations, 3 are in fact fixing the components $\phi^{0\mu}$ (if we assume that the non-degeneracy condition holds), and the remaining 3 have to be imposed as secondary constraints. Projecting onto the vector $B_\mu$, these secondary constraints can be written as
\be\nonumber
\Psi_\mu\equiv\eps^{ab}\,\Tr(F_{ab},B_\mu)\approx0.
\ee

\subsubsection*{{c. Evolution of the simplicity constraints $C_{\mu \nu}$}}

\noindent To study the evolution of the simplicity constraint, it is convenient to compute separately the Poisson bracket of its various components with the Hamiltonian. Requiring that the constraints $\mathcal{C}_{00}$ and $\mathcal{C}_{0a}$ be preserved under time evolution leads to
\begin{subequations}\label{fixation}
\ba
&&\dot{\mathcal{C}}_{00}=-2\Tr(\mu_0,\star B_0)\approx0,\\
&&\dot{\mathcal{C}}_{0a}=\Tr(\star B_0,\mathcal{D}_aB_0)+\Tr(\star B_0,[B_a,\omega_0])-\Tr(\mu_0,\star B_a)\approx0.
\ea
\end{subequations}
These 3 equations determine 3 of the 6 components of the multiplier $\mu_0$, and imply no secondary constraints on the dynamical variables of the phase space. Finally, a direct computation shows that the requirement that $\dot{\mathcal{C}}_{ab}\approx0$ leads to the 3 secondary constraints
\be\label{SSC}
\Phi_{ab}\equiv\Tr(\mathcal{D}_aB_0,\star B_b)+\Tr(\mathcal{D}_bB_0,\star B_a)\approx0.
\ee
The Dirac algorithm stops here, and there are no tertiary constraints.

\subsubsection*{{d. Summary of the analysis of the primary constraints}}

\noindent To summarize, we have a theory with 18 primary constraints consisting of 6 constraints $\pi_0$, 6 constraints $T_{ab}$, and 6 constraints $\mathcal{C}_{\mu\nu}$. They generate the 6 secondary constraints comprising the 3 constraints $\Psi_\mu$ and the 3 constraints $\Phi_{ab}$.

\subsubsection{\textbf{First and second class constraints}}

\noindent In the previous subsection, we have derived the primary and secondary constraints of the Plebanski theory. We are now going to split them between first class and second class constraints.

\subsubsection*{a. The first class constraints}

\noindent First, it is easy to see that the constraints $G$ are first class. Indeed, they commute with all the other constraints since they generate the infinitesimal $\SO(4)$ gauge symmetries. 

The analysis of equation (\ref{fixation}) suggests that amongst the 6 constraints $\pi_0\approx0$, 3 are first class and 3 are second class. To see that this is indeed the case, let us decompose the set of constraints $\pi_0\approx0$ into
\be\nonumber
\mathcal{K}_\mu\equiv\Tr(\pi_0,B_\mu),\qquad
\widetilde{\mathcal{K}}_\mu\equiv\Tr(\pi_0,\star B_\mu).
\ee
If the $B$ field does not belong to the degenerate sector, the previous set of constraints is equivalent to the requirement that $\pi_0 \simeq 0$. Furthermore, a direct computation (see the algebra of constraints in appendix \ref{constraints algebra}) shows that the constraints $\mathcal{K}_\mu$ are first class. They are associated with the 3 components of $\mu_0$ that remain undetermined after taking (\ref{fixation}) into account.

Finally, the last first class constraints are given by a new set $\widetilde{\Psi}_\mu$, which is obtained by adding to $\Psi_\mu$ a linear combination of constraints. In particular, $\widetilde{\Psi}_0$ is defined as being equal (up to a factor 2) to the Hamiltonian density (\ref{hamiltonian}):
\be\nonumber
\widetilde{\Psi}_0=
 {\Psi}_0
+2\eps^{ab}\,\Tr(\omega_0,T_{ab})
+\phi^{\mu\nu}\,\Tr(B_\mu,\star B_\nu)
+2 \Tr(\mu_0,\pi_0),
\ee
where the values of the Lagrange multipliers $\omega_0$, $\phi^{\mu \nu}$ and $\mu_0$, are those determined by the constraint analysis. When the Lagrange multipliers are not fixed by the Hamiltonian analysis, it is possible to fix them to the value zero in the expression for $\widetilde{\Psi}_0$. Concerning the constraints $\widetilde{\Psi}_a$, a direct calculation shows that they are given by
\ba
\widetilde{\Psi}_a&= &
\Psi_a + 2\eps^{bc}\,\Tr(\omega_a,T_{bc})-\Tr(\pi_0,\partial_aB_0)\nonumber\\
&=&2\eps^{bc}\left(\Tr(B_a,\partial_b\omega_c)+\Tr(\omega_a,\partial_bB_c)\right)-\Tr(\pi_0,\partial_aB_0),\nonumber
\ea
and that they generate as expected space diffeomorphisms on the phase
space variables:
\be\nonumber
\lbrace\widetilde{\Psi}(\xi),\omega_a\rbrace=-\mathcal{L}_\xi\omega_a,\qquad\lbrace\widetilde{\Psi}(\xi),B_\mu\rbrace=-\mathcal{L}_\xi B_\mu,
\ee
where $\xi$ is a vector field on $\Sigma$, and $\widetilde{\Psi}(\xi)$ denotes the smearing of the $\widetilde{\Psi}_a$ with $\xi^a$.

\subsubsection*{b. The second class constraints}

\noindent The remaining constraints, $\mathcal{C}_{\mu\nu}$, $\Phi_{ab}$ and $\widetilde{K}_\mu$, are the second class constraints of the theory. In order to prove this, it is possible to compute the (square 12-dimensional) Dirac matrix $\Delta$, whose elements are given by Poisson brackets between the various (candidate) second class constraints, and show that its determinant is non-vanishing. This is indeed the case, since the Dirac matrix is given by
\be\nonumber
\begin{tabular}{c|cccc}
$\Delta_{\alpha\beta}$	& \quad $\mathcal{C}_{0\mu}$ \quad	& \quad$\mathcal{C}_{ab}$\quad 	& \quad$\widetilde{\mathcal{K}}_\mu$\quad 	& \quad$\Phi_{ab}$\quad 	\\
  \hline
$\mathcal{C}_{0\nu}$		& 		$=0$					&		$=0$					&			$V_{\mu\nu}$					&		$=0$			\\ 
$\mathcal{C}_{cd}$ 		& 		$=0$					&		$=0$					&			$=0$							&		$M_{ab,cd}$		\\  
$\widetilde{\mathcal{K}}_\nu$	& 	$-V_{\nu\mu}$			&		$=0$					&			$\approx0$						&		$N_{ab,\nu}$		\\  
$\Phi_{cd}$				& 		$=0$					&		$-M_{cd,ab}$				&			$-N_{cd,\mu}$					&		$P_{ab,cd}$		\\  
\end{tabular}
\ee
and it satisfies clearly $\det(\Delta)\neq0$. Here the elements of the matrix $V$ are given by
\be\nonumber
V_{\mu\nu}\equiv\lb\widetilde{\mathcal{K}}_\mu,\mathcal{C}_{0\nu}\rb=\lb\Tr(\pi_0,\star B_\mu),\Tr(B_0,\star B_\nu)\rb=\p g_{\mu\nu}+\m g_{\mu\nu}=2g_{\mu\nu},
\ee
and we have $|\det(V)|\propto\mathcal{V}^2$, where $\mathcal{V}\equiv\sqrt{\det(g_{\mu\nu})}$ denotes the three-dimensional volume. The matrix $M$ has elements determined by the Poisson brackets 
\ba
M_{ab,cd}\equiv\lb\Phi_{ab},\mathcal{C}_{cd}\rb&=&\eps_{ca}\Tr([B_0,B_b],B_d)+\eps_{da}\Tr([B_0,B_b],B_c)\nonumber\\
&&+\eps_{cb}\Tr([B_0,B_a],B_d)+\eps_{db}\Tr([B_0,B_a],B_c),\nonumber
\ea
and satisfies $|\det(M)|\propto\mathcal{V}^3$. The explicit form of the matrices $N_{ab,\nu}\equiv\lb\Phi_{ab},\widetilde{\mathcal{K}}_\nu\rb$ and $P_{ab,cb}\equiv\lb\Phi_{ab},\Phi_{cd}\rb$ is not important in order to prove the invertibility of the Dirac matrix, and it is in fact easy to see that its determinant is given by\footnote{The exact proportionality coefficient is $c=2^{28}3^6$, since $|\det(V)|=2^3\mathcal{V}^2$, and $|\det(M)|=2^{11}3^3\mathcal{V}^3$.}
\be\nonumber
\det(\Delta)= \big(\det(V) \det(M)\big)^2 =c\mathcal{V}^{10}.
\ee

\subsubsection{\textbf{Summary}}

\noindent To summarize, the first class constraints of the system are given by $\mathcal{K_\mu}$, $\widetilde{\Psi}_\mu$ and $G_{ab}$, and the constraints $\mathcal{C}_{\mu\nu}$, $\Phi_{ab}$ and $\widetilde{K}_\mu$, are of second class. Therefore, we have 36 phase space variables $B^{IJ}_\mu$ and $\omega^{IJ}_\mu$ that are subject to 12 first class constraints (generating gauge symmetries) and 12 second class constraints, and the usual counting shows that there are no phase space degrees of freedom. This is of course to be expected in three-dimensional gravity.

As usual, the first class constraints of the theory are the infinitesimal generators of the gauge symmetries. The constraint $G$ is the generator of the internal $\SO(4)$ symmetries. The 3 primary constraints $\mathcal{K}_\mu$ appear because we have treated the lapse and the shift encoded in $B$ as dynamical variables. The constraints $\mathcal{K}_0$ and $\mathcal{K}_a$ therefore encode the vanishing of the momenta $\pi_N$ and $\pi_{N^a}$ conjugated to the lapse and the shift. The remaining constraints, $\widetilde{\Psi}_0$ and $\widetilde{\Psi}_a$, are related to the scalar and vector constraints, i.e. to the spacetime diffeomorphisms.

Let us conclude this canonical analysis with an important remark concerning the second class constraint. The 3 constraints $\widetilde{\mathcal{K}}_\mu$ constrain 3 out of the 6 components of the variables $\pi_0^{IJ}$, which are the conjugate momenta to $B_0^{IJ}$. As a consequence, only 3 components of $\pi_0^{IJ}$ are left, and they correspond to the momenta $\pi_N$ and $\pi_{N^a}$ associated with the lapse and shift variables. Moreover, the lapse and the shift are the 3 independent components of $B_0^{IJ}$ that are left after imposing the 3 components $\mathcal{C}_{0\mu}$ of the simplicity constraints. The spatial part $\mathcal{C}_{ab}$ of the simplicity constraints ensures that the components $B^{IJ}_a$ can be expressed in terms of the zero-form $\chi^I$ and the one-form $e^I_a$. Finally, the meaning of the 3 constraints $\Phi_{ab}$ is that 3 of the components of the connection $\omega^{IJ}_a$ can be expressed in terms of $e^I_a$ and $\chi^I$. It is important to stress that these are secondary constraints which have been obtained from the requirement that $\mathcal{C}_{ab}$ be preserved in time. In other words, the usual simplicity constraints are of second class because there are in fact secondary constraints which do not commute with the primary ones. This point will be very important in the spin foam quantization that we perform in the next section. We are going to show the importance of imposing the secondary second class constraints in the spin foam models.

\section{Spin foam quantization}
\label{sec2}

\noindent In this section, we study the quantization of the three-dimensional Plebanski theory. From the classical analysis of the previous section, it should be clear that the action (\ref{3d plebanski}) is equivalent in the gravitational sector to that of first order gravity, whose spin foam quantization is the Ponzano-Regge model. We can now try to reproduce the strategy of four-dimensional spin foam models, and see if the spin foam quantization of the full Plebanski theory leads to the Ponzano-Regge model as well. This is a well-posed question which will enable us to clarify the issue of imposing the simplicity constraints at the quantum level.

\subsection{The Ponzano-Regge model}

\noindent We have seen in the previous section that when the simplicity constraints are imposed at the classical level, the gravitational sector of solutions to the Plebanski action (\ref{3d plebanski}) is equivalent to the $\SU(2)$ Palatini action. It is therefore possible to compute explicitly the simplicial path integral of this theory, and we know that it leads to the Ponzano-Regge state sum model, whose vertex amplitude is given by an $\SU(2)$ $6j$ symbol. We recall some basic facts about the derivation of this result, and also introduce the notations that we will use in the next subsections.

The partition function of three-dimensional Riemannian gravity is formally given by
\be\nonumber
\mathcal{Z}_\text{grav}=\int\de[e]\de[\omega]\exp\left(i\int_\mathcal{M}\de^3x\,\eps^{\mu\nu\rho}\,\tr(e_\mu,F_{\nu\rho})\right),
\ee
where $\tr(\cdot\,,\cdot)$ denotes a trace in the Lie algebra $\su(2)$. To compute the simplicial path integral, we introduce a simplicial decomposition $\Delta$ of the spacetime manifold $\mathcal{M}$, along with its dual two-complex $\Delta^*$ \cite{rourke-sanderson}. This latter is consisting of vertices $v$ (dual to tetrahedra $\tau\in\Delta$), edges $e$ (dual to triangles $t\in\Delta$) and faces $f$ (dual to links $\ell\in\Delta$). Since the triad $e^i_\mu$ is a one-form, it is natural to integrate it along the one-cells (links) of $\Delta$ to obtain $\su(2)$-valued elements $X_f$. The connection is discretized by computing its holonomy $h_e$ along the edges $e$ of $\Delta^*$. We can then discretize the curvature by taking the product of holonomies along the edges lying on the boundary of a face $f\in\Delta^*$ to form
\be\nonumber
h_f=\prod_{e\subset f}h_e.
\ee
With these variables, the discretized version of the action (\ref{gravity action}) can be written as
\be\nonumber
S_\text{BF}[X_f,h_f]=\sum_{f\in\Delta^*}\tr(X_f,h_f),
\ee
and the partition function becomes
\be\label{PRintermediaire}
\mathcal{Z}_\text{grav}
=\left(\prod_{f\in\Delta^*}
\int_{\su(2)}\de X_f\right)\left(\prod_{\vphantom{f}e\in\Delta^*}
\int_{\SU(2)}\de h_e\right)\exp\left(i\sum_{f\in\Delta^*}\tr(X_f,h_f)\right),
\ee
where $\de X_f$ is the Lebesgue measure on $\su(2)\simeq\mathbb{R}^3$, and $\de h_e$ the Haar measure on $\SU(2)$. It is now possible to perform the integral over $X_f$ to obtain
\be\nonumber
\mathcal{Z}_\text{grav}
=\left(\prod_{\vphantom{f}e\in\Delta^*}\int_{\SU(2)}\de h_e\right)\prod_{f\in\Delta^*}\delta_{\SU(2)}(h_f),
\ee
where the delta distribution over $\SU(2)$ imposes the flatness of the connection. Using the Peter-Weyl decomposition, we can write
\ba
\mathcal{Z}_\text{grav}&=&\left(\prod_{\vphantom{f}e\in\Delta^*}
\int_{\SU(2)}\de h_e\right)\prod_{f\in\Delta^*}\sum_{\{j\}\rightarrow f}(2j_f+1)\chi_{j_f}(h_f)\nonumber\\
&=&\sum_{\{j\}\rightarrow\{f\}}\prod_{f\in\Delta^*}(2j_f+1)
\left(\prod_{\vphantom{f}e\in\Delta^*}\int_{\SU(2)}\de h_e\right)
\prod_{f\in\Delta^*}\chi_{j_f}\left(\prod_{e\subset f}h_e\right),\label{pf}
\ea
where the sum is taken over all the possible $\SU(2)$ representations $j$
labeling the set of faces $f\in\Delta^*$. For arbitrary cellular decompositions $\Delta$,
let us call $n$ the number of faces meeting at every edge $e\in\Delta^*$.
In (\ref{pf}) we will have an integral over $h_e$ of $n$ products
of representation matrices for the group element $h_e$.
We can therefore use the fact that
\be\label{group integral}
\int\de h_e\,\mathbf{D}^{(j_1)}(h_e)\dots\mathbf{D}^{(j_n)}(h_e)=\sum_{i_e}i_ei_e^*
\ee
projects onto the space $\text{Inv}_{\SU(2)}\left(\mathcal{H}^{(j_1)}_{\SU(2)}\otimes\dots\otimes\mathcal{H}^{(j_n)}_{\SU(2)}\right)$
of intertwiners between the $n$ representations coloring the faces meeting at the edge $e$.
Then, all the intertwiners meeting at a vertex $v$ can be contracted
to define a vertex amplitude $A_v(j_{f\supset v},i_{e\supset v})$.
Finally, the partition function can be written as a sum over spin foam amplitudes
\be\nonumber
\mathcal{Z}_\text{grav}=\sum_{\{j\}\rightarrow\{f\}}\sum_{\{i\}\rightarrow\{e\}}
\prod_{f\in\Delta^*}(2j_f+1)\prod_{v\in\Delta^*}A_v(j_{f\supset v},i_{e\supset v}).
\ee
To clarify the meaning of this formula, let us assume that the cellular decomposition
$\Delta$ is simplicial. In this case, vertices in $\Delta^*$
are four-valent, while edges are three-valent. The vertex amplitude
is therefore given by a contraction of four three-valent intertwiners,
which is the so-called $6j$ symbol. The final result of this calculation is
\be\label{3d partition function}
\mathcal{Z}_\text{grav}
=\sum_{\{j\}\rightarrow\{f\}}\prod_{f\in\Delta^*}(2j_f+1)\prod_{v\in\Delta^*}\{6j\}
%=\sum_{\{j\}\rightarrow\{f\}}\prod_{f\in\Delta^*}(2j_f+1)\prod_{v\in\Delta^*}
%\left\{\begin{array}{ccc}
%         j_1 & j_2 & j_3\\
%         j_4 & j_5 & j_6\end{array}\right\},
\ee
where the labels $j_1,\dots j_6$ are associated to the six faces dual to the links $\ell$
a tetrahedron $\tau\in\Delta$. Notice that the sum over intertwiners
has now disappeared since there is only a unique (up to normalization)
three-valent intertwiner. In other words, the intertwiner space
\be\nonumber
\text{Inv}_{\SU(2)}\left(\mathcal{H}^{(j_1)}_{\SU(2)}\otimes\mathcal{H}^{(j_2)}_{\SU(2)}\otimes\mathcal{H}^{(j_3)}_{\SU(2)}\right)
\ee
is one-dimensional when $(j_1,j_2,j_3)$ satisfy the triangular inequalities, and zero-dimensional otherwise.

It is convenient at this point to introduce the diagrammatic notations \cite{oeckl}, in which lines are used to represent unitary irreducible representations, and boxes to represent integrals over the group defining a projector on the intertwiner space following (\ref{group integral}). With this notation, the partition function defining the Ponzano-Regge model can be represented as follows:
\be\nonumber
\mathcal{Z}_\text{grav}=\sum_{\{j\}\rightarrow\{f\}}\prod_{f\in\Delta^*}(2j_f+1)
\vcenter{\hbox{\includegraphics[scale=0.3]{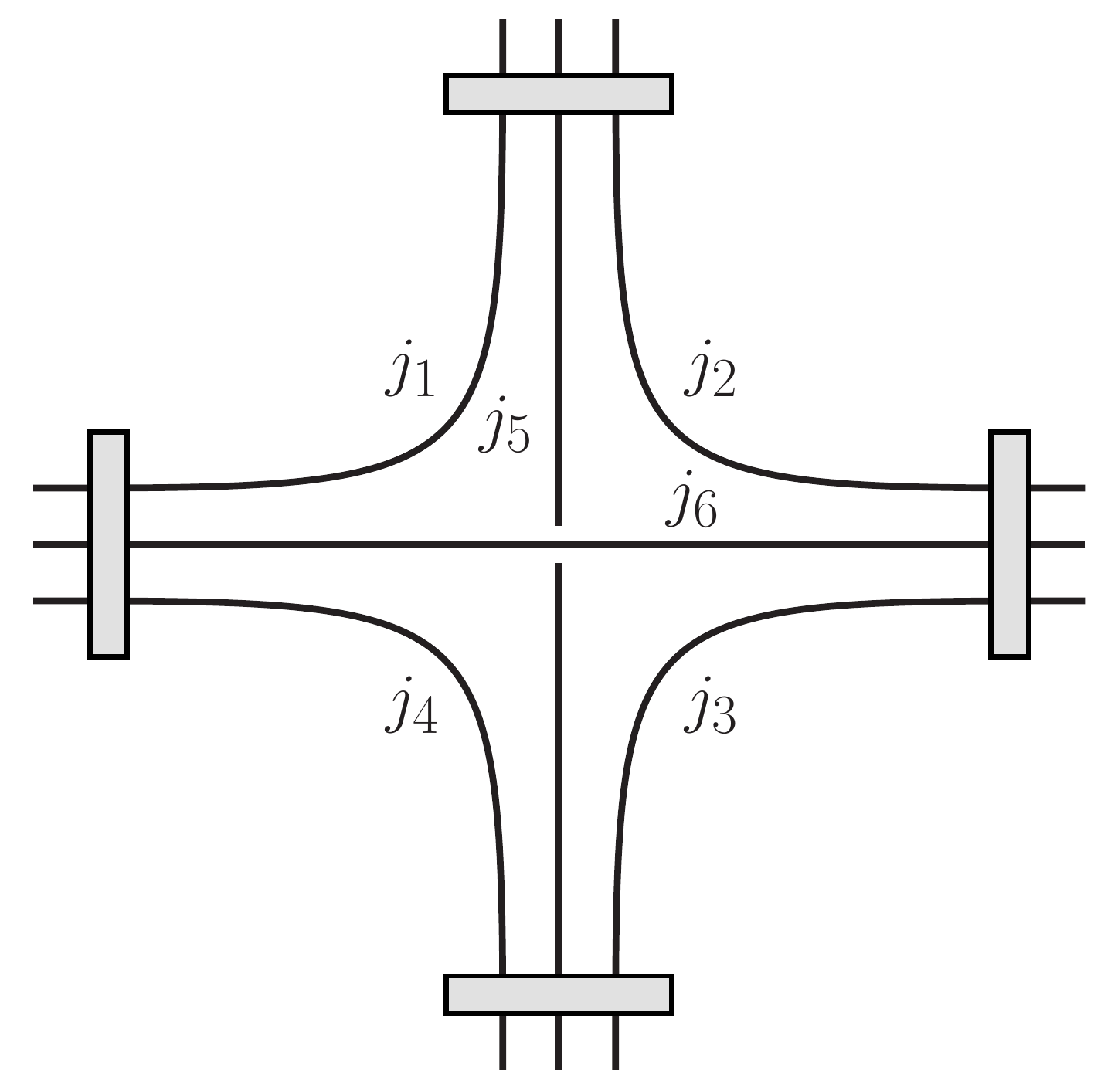}}}\quad.
\ee
Once the integrations are performed, we obtain a contraction a four three-valent intertwiners for each vertex dual to a tetrahedron, and the partition function becomes (\ref{3d partition function}).

\subsection{Quantization of the $\boldsymbol{\SO(4)}$ $\boldsymbol{BF}$ sector}

\noindent Spin foam models in four dimensions take as a starting point the four-dimensional Plebanski action, and are derived by first quantizing the topological part of the action, and then imposing the simplicity constraints at the quantum level.

Here we follow the same procedure, and start by quantizing the $\SO(4)$ $BF$ part of the Plebanski action (\ref{3d plebanski}). This can be done very easily along the lines of the standard construction introduced in the previous subsection, with the only difference that $\SU(2)$ is now replaced by $\SO(4)$. Therefore, the $B$ field is discretized by assigning an $\so(4)$ element $B_f$ to each face $f\in\Delta^*$ dual to a one-cell in $\Delta$. Once we perform the integration over $B_f$ in the simplicial path integral, we are left with
\be\nonumber
\mathcal{Z}_\text{BF}
=\left(\prod_{\vphantom{f}e\in\Delta^*}\int_{\SO(4)}\de h_e\right)\prod_{f\in\Delta^*}\delta_{\SO(4)}(h_f),
\ee
where now we have a product of delta distributions on the group $\SO(4)$, which can again be evaluated with the Peter-Weyl decomposition. Using the fact that $\SO(4)\simeq\SU(2)\times\SU(2)/\text{Z}_2$, it is possible to write an irreducible representation $\rho$ of $\SO(4)$ as a tensor product $\rho=j^+\otimes j^-$ of two irreducible representations of $\SU(2)$. The discretized partition function for the three-dimensional $\SO(4)$ $BF$ theory then takes the form
\be\nonumber
\mathcal{Z}_\text{BF}
=\sum_{\{j^+,j^-\}\rightarrow\{f\}}\prod_{f\in\Delta^*}(2j^+_f+1)(2j^-_f+1)\prod_{v\in\Delta^*}\{6j^+\}\{6j^-\}.
\ee
Now, we have to implement the simplicity constraints in this state sum in order to ensure that it describes quantum three dimensional gravity. Indeed, since at the classical level the gravitational sector of the Plebanski action corresponds to the action for gravity, whose spin foam quantization corresponds to (\ref{3d partition function}), we expect that a proper imposition of the simplicity constraints will lead to the same result.

\subsection{Imposition of the simplicity constraints}

\noindent In the simplicial path integral that we have written for the three-dimensional $\SO(4)$ $BF$ theory, the one-forms $B^{IJ}_\mu$ have been discretized by assigning an element $B^{IJ}_f\in\so(4)$ to the faces of $\Delta^*$, which are dual to the links $\ell\in\Delta$ defining the boundary $\partial t$ of the two-cells (triangles) $t\in\Delta$. These vectors in fact determine the geometry of the triangles, and they satisfy the closure constraint
\be\nonumber
\sum_{f\in\partial t}B^{IJ}_f=0.
\ee
This is just the discrete analogue of the continuous Gauss law which ensures gauge invariance under the action of $\SO(4)$. Additionally, the discretized $B$ field is required to satisfy the discretized version of the simplicity constraints $\mathcal{C}$, which are given by\footnote{We do not mention the additional volume constraints, since they are automatically implemented when the closure, diagonal and off-diagonal simplicity constraints are satisfied.}
\ba
\text{diagonal simplicity:}\qquad&&\eps_{IJKL}B^{IJ}_fB^{KL}_f\approx0,\qquad\forall\,f\in\partial t,\label{diagonal}\\
\text{off-diagonal simplicity:}\qquad&&\eps_{IJKL}B^{IJ}_fB^{KL}_{f'}\approx0,\qquad\forall\,f,f'\in\partial t.\label{off-diagonal}
\ea
These simplicity constraints do not distinguish between the topological and gravitational sectors, since they are left unchanged if we change $B$ for $\star B$.

Just like in the four-dimensional theory \cite{FK,alexandrov3}, the simplicity constraints can be linearized, and are equivalent to the requirement that
\be\label{linear simplicity}
\chi_IB^{IJ}_f=0,\qquad\forall\,f\in\partial t.
\ee
Indeed, in the gravitational sector where we have $B^{IJ}_\mu=\eps^{IJ}_{~~KL}\chi^Ke^L_\mu$, it is clear that the linear simplicity constraint is satisfied. This linearized version has therefore the advantage of selecting the gravitational sector.

The quantization of the theory is based on the symplectic structure of the topological part of the action. In the $BF$ action, the variable $B$ is canonically conjugated to the connection $\omega$, and the quantization rule is therefore simply to identify the discretized field $B_f$ with the generators $\hat{B}_f\equiv J_f$ of the Lie algebra $\so(4)$. The next step is to impose the simplicity constraints on these generators following the different spin foam models that have been introduced in the literature.

The imposition of the constraints is done before the integration over the connection components defining the intertwiners. Schematically, we can represent this imposition of the constraints as a white box acting on the group generators between two neighboring triangles. Our task is therefore to implement the simplicity constraints in such a way that the partition function reduces to
\be\label{constrained path integral}
\sum_{\{j^+,j^-\}\rightarrow\{f\}}\prod_{f\in\Delta^*}(2j^+_f+1)(2j^-_f+1)
\vcenter{\hbox{\includegraphics[scale=0.3]{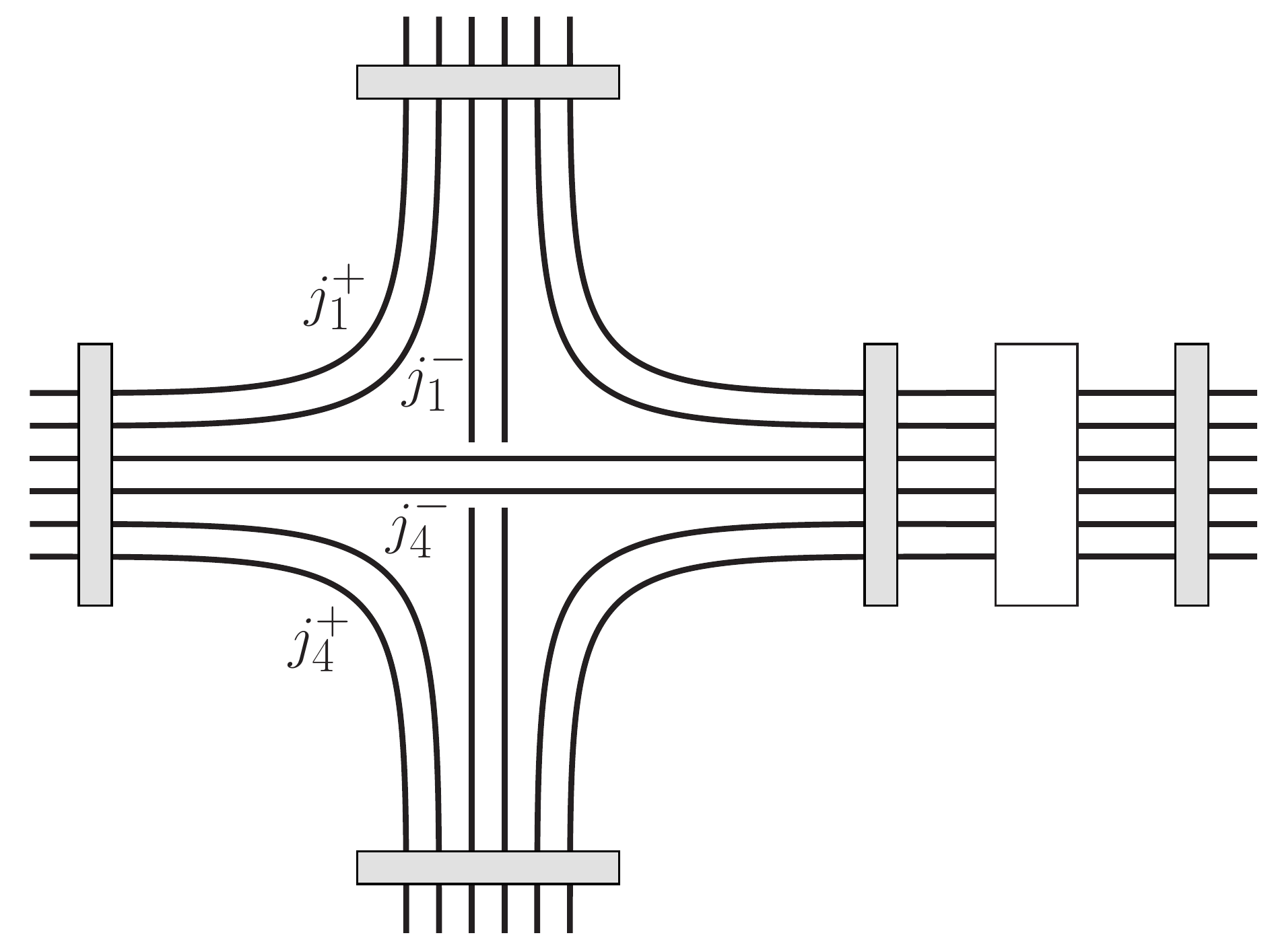}}}
\quad\overset{?}{=}\quad
\mathcal{Z}_\text{grav}.
\ee
Here we have represented the implementation of the constraints via $\hat{\mathcal{C}}$ on one triangle only for the sake of clarity. More explicitly the meaning of this graphical notation is that we have an integral over the $\SO(4)$ holonomies, together with a yet-to-be defined implementation of the simplicity constraints $\hat{\mathcal{C}}$. In other words, we have
\be\label{operator}
\vcenter{\hbox{\includegraphics[scale=0.3]{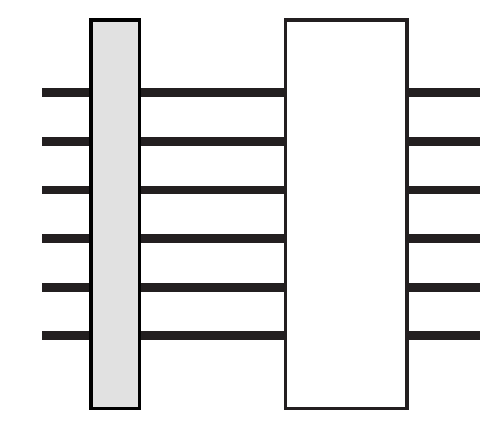}}}
\quad\equiv\quad
\int\de h\,\delta(\hat{\mathcal{C}}),
\ee
and the implementation of the simplicity constraints is done before the integration over the holonomies that defines the intertwiners leaving on the edges.

\subsubsection{\textbf{The BC prescription}}

\noindent We are going to start by solving the diagonal simplicity constraint. It is known that this constraint is equivalent to the requirement that the pseudo-scalar quadratic Casimir operator of $\so(4)$ defined by $C_2\equiv\Tr(\star J,J)$ vanishes, which constrains the representations to be simple. Indeed, in terms of the self-dual and anti self-dual generators, (\ref{diagonal}) takes the form
\be\nonumber
\Tr(\star J_f,J_f)=\Tr\big((\p J_f-\m J_f),(\p J_f+\m J_f)\big)=L_f\cdot K_f=0,
\ee
which implies the restriction $j^+_f=j^-_f\equiv j_f$ to simple representations of $\SO(4)$ \cite{BC,FK-simple}. The off-diagonal simplicity constraint involves two different faces $f$ and $f'$ dual to links $\ell$ and $\ell'$ belonging to the same triangle. They are constraints on the intertwiner states of the quantum theory, and if we impose them strongly we obtain the unique Barrett-Crane intertwiner \cite{BC}, which selects only the subspace $j_f=0$ of the decomposition
\be\label{SO(4) decomposition}
\mathcal{H}^{(j_f,j_f)}_{\SO(4)}=\bigoplus_{j_f=0}^{2j_f}\mathcal{H}_{\SU(2)}^{(j_f)}.
\ee
Indeed, it is clear that the strong imposition of the constraint $C_1\equiv\Tr(J_f,J_f)=2j_f(j_f+1)$ imposes $j_f=0$. Therefore, the operator implementing the simplicity constraint in the BC model acts like
\be\label{operatorBC}
\mathcal{P}_\text{BC}(j_1,j_2,j_3)
\quad\equiv\quad
\vcenter{\hbox{\includegraphics[scale=0.3]{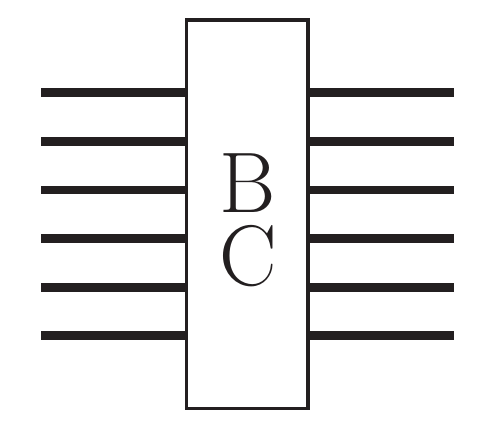}}}
\quad=\quad
\vcenter{\hbox{\includegraphics[scale=0.3]{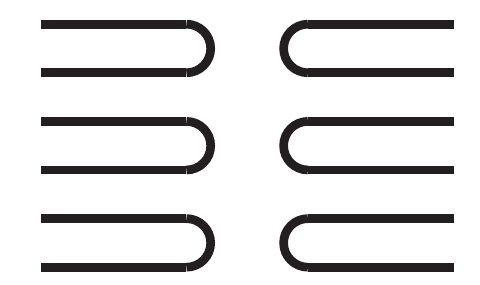}}}\quad.
\ee
As a result, if we use this operator in the path integral (\ref{constrained path integral}), we obtain the state sum
\ba
\mathcal{Z}_\text{BC}
&=&\sum_{\{j\}\rightarrow\{f\}}\prod_{f\in\Delta^*}(2j_f+1)^2\vcenter{\hbox{\includegraphics[scale=0.3]{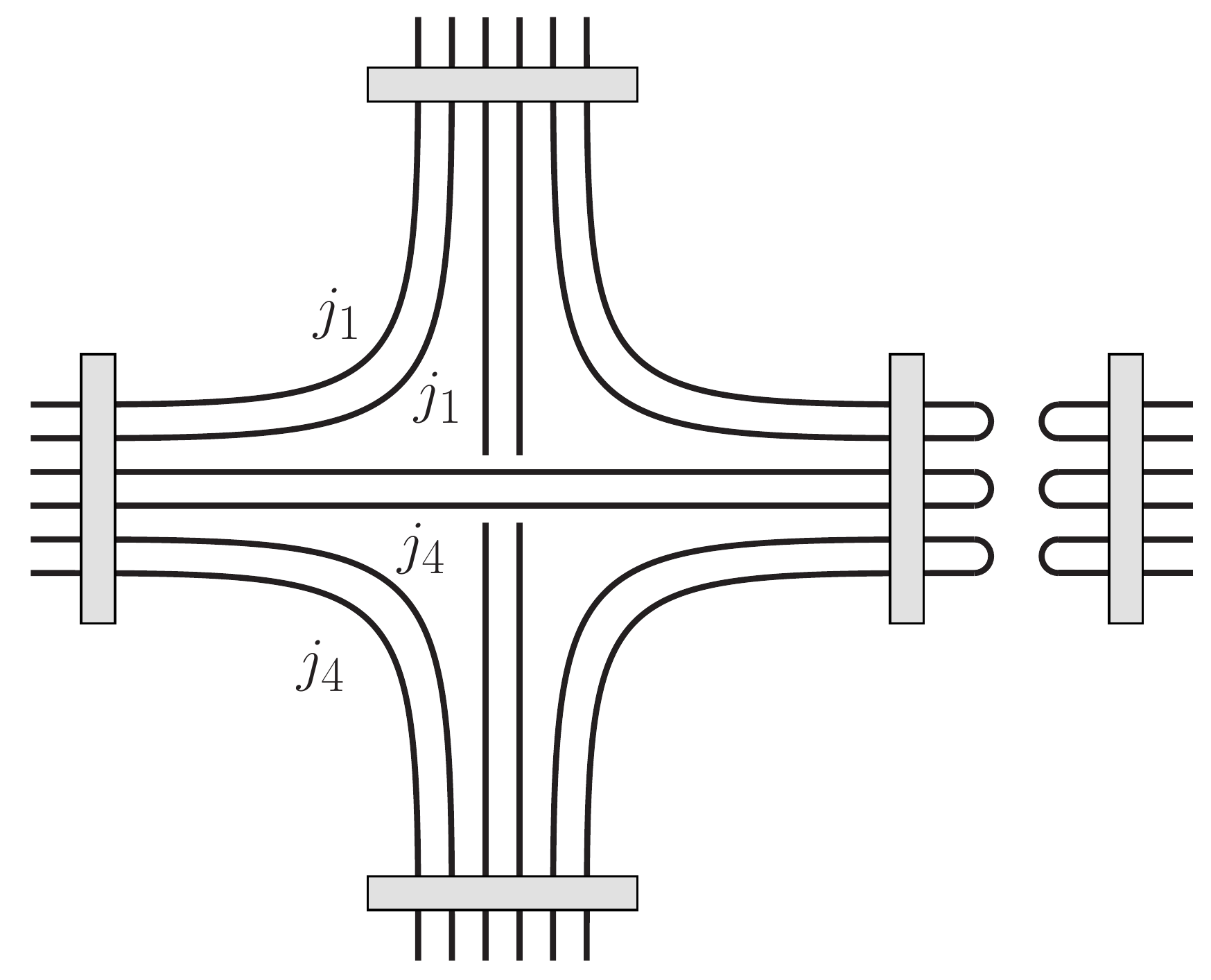}}}\nonumber\\
&=&\sum_{\{j\}\rightarrow\{f\}}\prod_{f\in\Delta^*}(2j_f+1)^2\prod_{v\in\Delta^*}\{6j\}^2.\nonumber
\ea
Notice that the BC operator defined in (\ref{operatorBC}) is not a projector. However, it can be renormalized to give the genuine projector
\be\nonumber
\widetilde{\mathcal{P}}_\text{BC}(j_1,j_2,j_3)\equiv\f{1}{d_{j_1}d_{j_2}d_{j_3}}\mathcal{P}_\text{BC}(j_1,j_2,j_3),
\ee
where $d_{j_i}=2j_i+1$ denotes the dimension of the spin-$j$ representation. This modification only affects the weight associated to the faces, and not the vertex amplitude.

We see from this construction that the BC prescription, in which the simplicity constraints are imposed strongly, does not lead to the proper vertex amplitude. The next step is to check wether a weak imposition of the constraints can cure this discrepancy.

\subsubsection{\textbf{The EPRL prescription}}

\noindent Now we want to test the EPRL imposition of the simplicity constraints. The diagonal simplicity constraint can again be imposed strongly, and leads to the restriction $j_f^+=j_f^-\equiv j_f$ to simple representations. The idea is then to impose the off-diagonal constraint weakly. For this, we use the linear constraint (\ref{linear simplicity}), and notice that in the time gauge it becomes
\be\nonumber
B^{0i}_f=K^i_f\approx0.
\ee
One way to impose this constraint weakly is to use the master constraint trick \cite{EPR}, and impose the vanishing of the square of the boost components $K^i_f$. From the expression of the quadratic Casimir operators of $\SO(4)$, we find that this amounts to representing $\hat{\mathcal{C}}$ by a projection and an integration over the highest weight representations, i.e. $j^+_f+j^-_f=2j_f$. The EPRL operator realizing this imposition of the simplicity constraints can be represented as
\be\label{operatorEPRL}
\mathcal{P}_\text{EPRL}(j_1,j_2,j_3)
\quad\equiv\quad
\vcenter{\hbox{\includegraphics[scale=0.3]{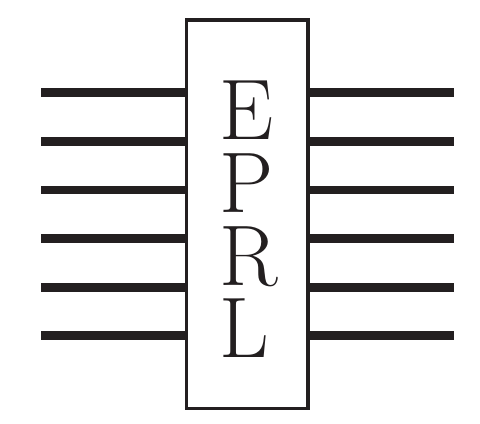}}}
\quad=\quad
\vcenter{\hbox{\includegraphics[scale=0.3]{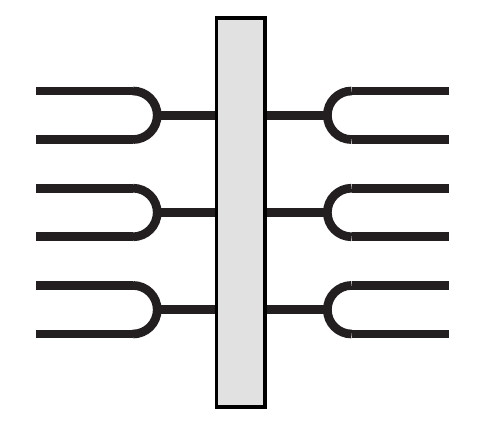}}}\quad,
\ee
where the integration is taken over the spin-$2j_f$ representations. With this implementation of the simplicity constraints, the partition function (\ref{constrained path integral}) becomes
\ba
\mathcal{Z}_\text{EPRL}
&=&\sum_{\{j\}\rightarrow\{f\}}\prod_{f\in\Delta^*}(2j_f+1)^2\vcenter{\hbox{\includegraphics[scale=0.3]{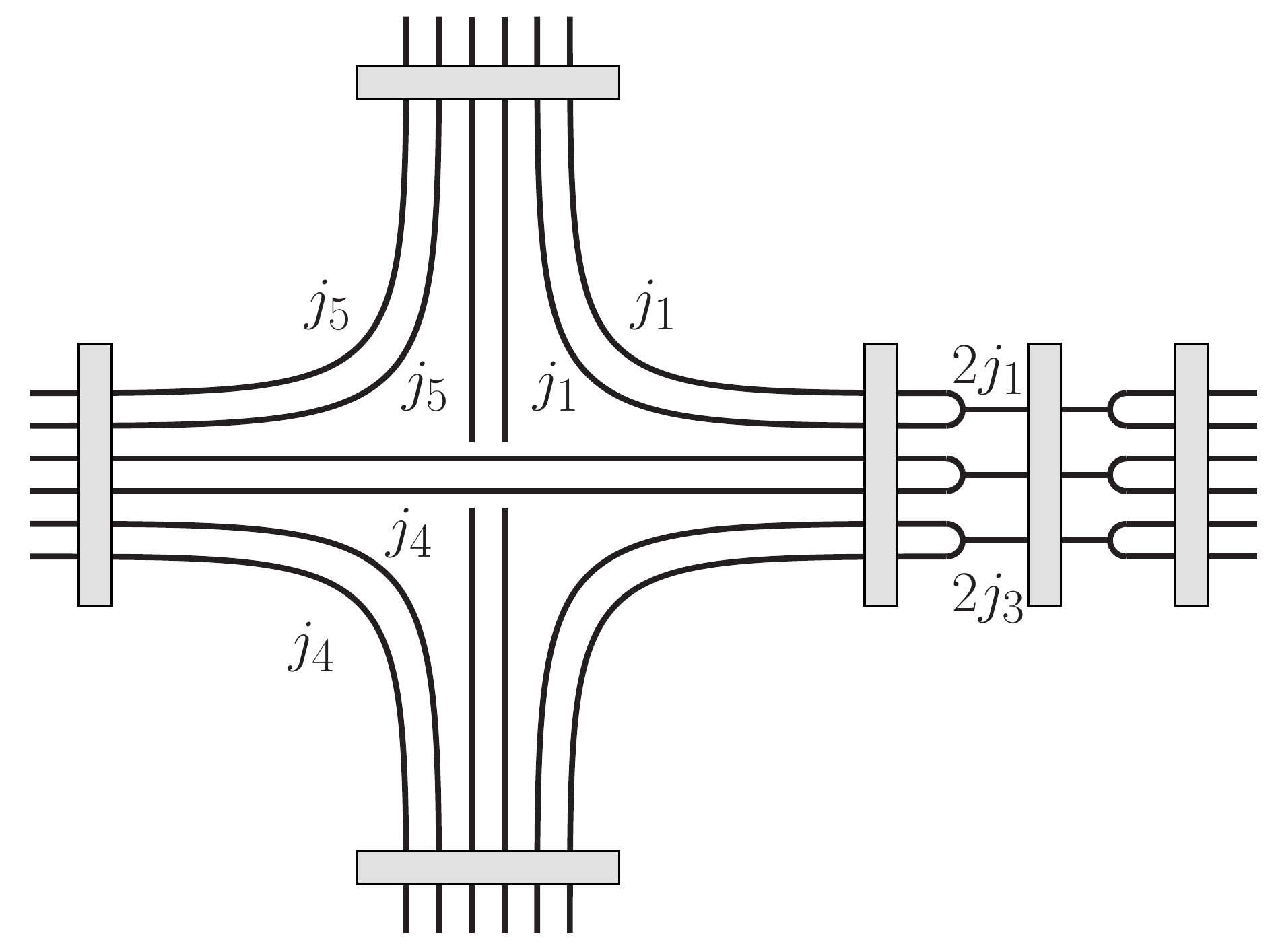}}}\nonumber\\
&=&\sum_{\{j\}\rightarrow\{f\}}\prod_{f\in\Delta^*}(2j_f+1)^2\prod_{v\in\Delta^*}\left(\{6j\}^2\prod_{\alpha=1}^4f_\alpha(j_{\alpha\beta})\right),\nonumber
\ea
with $\alpha,\beta=1,\dots,4$, the index $\alpha$ labeling the edges, and the pair $\alpha\beta$ labeling the faces. The fusion coefficient $f_\alpha(j_1,j_2,j_3)$ attached to each edge $e$ is given by the following $9j$ symbol:
\be\label{9j}
\left\{
\begin{array}{ccc}
j_1&j_2&j_3\\
j_1&j_2&j_3\\
2j_1&2j_2&2j_3
\end{array}\right\}
\quad=\quad\vcenter{\hbox{\includegraphics[scale=0.35]{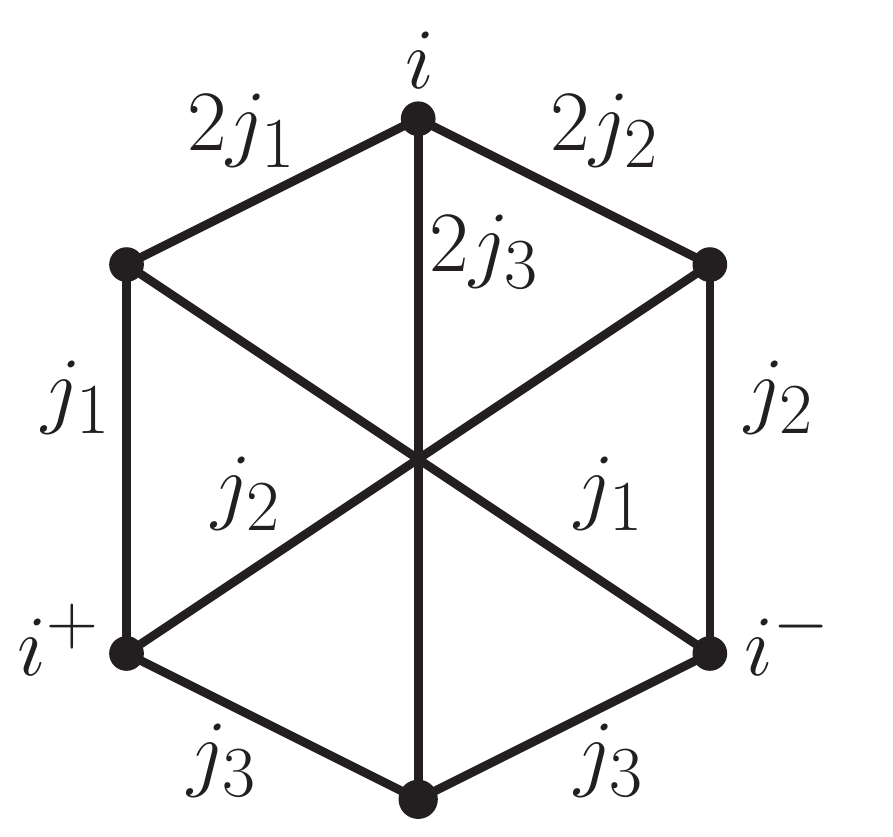}}}.
\ee
Notice that the dependance of the fusion coefficient on the intertwiners $i$, $i^+$ and $i^-$, drops out because these intertwiners are three-valent and therefore unique. This is however an artifact of the fact that we have chosen the cellular decomposition $\Delta$ of the spacetime manifold to be simplicial (each edge of $\Delta^*$ is therefore bounded by three faces). Another choice would have lead to intertwiners with higher valency.

\subsubsection{\textbf{The Warsaw modification of the EPRL prescription}}

\noindent Just like in the four-dimensional case, the operator (\ref{operatorEPRL}) that we have in the previous subsection to implement the simplicity constraints is not a projector, i.e.
\be\nonumber
\left(\vcenter{\hbox{\includegraphics[scale=0.3]{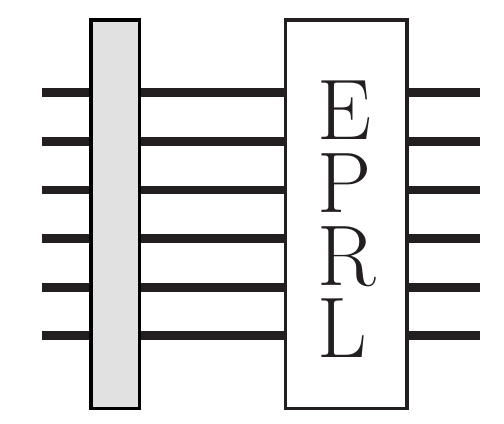}}}\right)^2
\quad\neq\quad
\vcenter{\hbox{\includegraphics[scale=0.3]{operatorEPRL.pdf}}}\quad.
\ee
This ``problem'' can be solved following the idea introduced in the four-dimensional context in \cite{KKL1,KKL2,BHKKL}, and defining a normalized operator
\be\nonumber
\vcenter{\hbox{\includegraphics[scale=0.3]{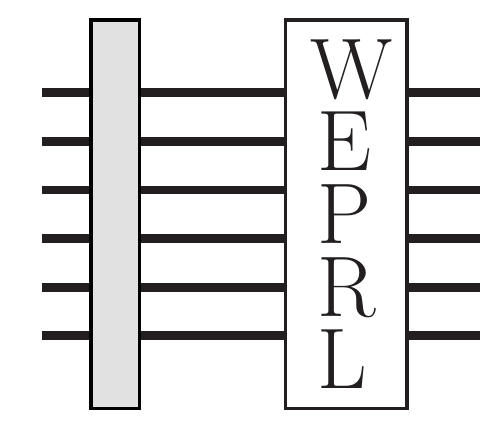}}}
\quad\equiv\quad
\f{1}{\{9j\}}\vcenter{\hbox{\includegraphics[scale=0.3]{operatorEPRL.pdf}}}\quad,
\ee
which is now a genuine projector. The $9j$ symbol appearing here is the one given by (\ref{9j}). This normalization factor is in some sense very trivial because we are dealing with a three-dimensional theory and the sum over the (three-valent) intertwiners has disappeared. As a result, if we use this projector to impose the simplicity constraints, the $9j$ in its definition compensates the fusion coefficient obtained from the EPRL operator, and we obtain that the vertices of the BC and Warsaw-EPRL models are the same.

Therefore, in the three-dimensional model at hand, and in the absence of Immirzi parameter, the prescription of defining a true projector to impose the simplicity constraints does not lead to the proper vertex amplitude either.

\subsection{Imposition of the secondary second class constraints}

\noindent In the previous subsection, we have tried to impose the simplicity constraints in the partition function for the $\SO(4)$ $BF$ theory in order to recover the spin foam quantization of the gravitational sector of the classical Plebanski theory (\ref{3d plebanski}), i.e. the Ponzano-Regge model. The partition functions $\mathcal{Z}_\text{BC}$ and $\mathcal{Z}_\text{EPRL}$ that we have obtained differ from that of the Ponzano-Regge model by the form of the face and vertex amplitudes. 

The form of the face amplitude is supposed to be related to the path integral measure. In the spin foam quantization (see for example equation (\ref{PRintermediaire})), the measures are taken to be the Haar measure for $\SU(2)$- or $\SO(4)$-valued elements, and the usual Lebesgue measure for Lie algebra-valued elements. However, it is known that the Plebanski action has a non-trivial path integral measure due to the presence of the various types of constraints \cite{BHNR,EHT,perez-bojowald}. It would be interesting to use the results derived during the canonical analysis of section \ref{sec2} (in particular, the determinant of the Dirac matrix) to compute the path integral measure of the Plebanski theory (\ref{3d plebanski}), and see if we can recover the face amplitude $(2j_f+1)$ of the Ponzano-Regge model. We plan to come back to this question in future work.

For the time being, let us focus on the vertex amplitude. To derive the spin foam models of the previous subsections, following the prescriptions of the BC and EPRL models, we have implemented the second class simplicity constraints $\mathcal{C}$. However, as long emphasized by Alexandrov \cite{revue,alexandrov1,alexandrov2,alexandrov3} for instance, the primary simplicity constraints $\mathcal{C}$ generate secondary constraints with whom they form a second class pair, and these secondary second class constraints should also be imposed in the derivation of the spin foam models. Indeed, we have seen in the canonical analysis of the Plebanski action in section \ref{sec:canonical} that the requirement that the primary simplicity constraint $\mathcal{C}_{\mu\nu}$ be preserved in time generates the secondary constraints $\Phi_{ab}$. These constraints are additional second class constraints that have to be implemented somehow.

Let us have a closer look at the constraints $\Phi_{ab}$ given by (\ref{SSC}). For the sake of clarity, we only focus on the first term, which can be written with the decomposition into self-dual and anti self-dual fields in the form
\ba
\Tr(\mathcal{D}_aB_0,\star B_b)&=&\Tr(\partial_aB_0,\star B_b)+\Tr([\omega_a,B_0],\star B_b)\nonumber\\
&=&\Tr(\partial_a\p B_0,\p B_b)-\Tr(\partial_a\m B_0,\m B_b)\nonumber\\
&&+\,\Tr([\p\omega_a,\p B_0],\p B_b)-\Tr([\m\omega_a,\m B_0],\m B_b).\nonumber
\ea
From equation (\ref{selfdual BG}), we see that in the time gauge and in the gravitational sector we have $\p B^i_\mu=\m B^i_\mu$. The previous equation therefore reduces to
\be\nonumber
\Tr(\mathcal{D}_aB_0,\star B_b)=\Tr([\p\omega_a-\m\omega_a,\p B_0],\p B_b),
\ee
and we have a similar relation for the symmetric term. It is therefore clear that the condition $\p\omega=\m\omega$ solves the constraint $\Phi_{ab}$ (in the gravitational sector and in the time gauge). To implement this additional constraint at the level of the partition function, let us first discretize the condition $\p\omega=\m\omega$, and write it in terms of holonomies as $\p h=\m h$. We can then replace the operator (\ref{operator}) by
\be\nonumber
\vcenter{\hbox{\includegraphics[scale=0.3]{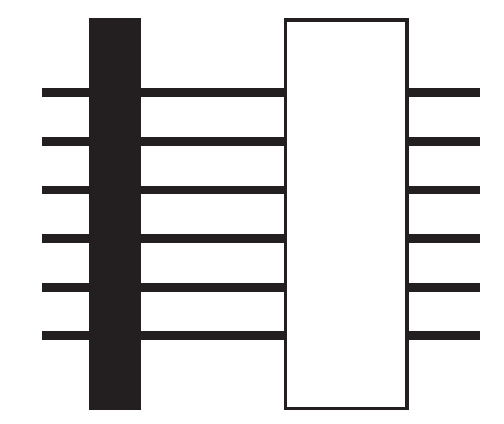}}}
\quad\equiv\quad
\int\de h\,\delta(\Phi)\delta(\hat{\mathcal{C}})\quad=\quad\int\de h\,\delta\left(\p h\m h^{-1}\right)\delta(\hat{\mathcal{C}}),
\ee
where the white box should implement the simplicity constraints $\mathcal{C}$, and the integration measure has been changed to take into account the imposition of the secondary second class constraints $\Phi$.

If we impose the simplicity constraints $\delta(\hat{\mathcal{C}})$ \`a la BC, i.e. by projecting first onto simple representations $j^+_f=j^-_f\equiv j_f$, and then considering only the lowest weight module in the decomposition (\ref{SO(4) decomposition}), we obtain a trivial vertex amplitude which is simply given by
\be\nonumber
A=d_{j_1}d_{j_2}d_{j_3}d_{j_4}d_{j_5}d_{j_6}.
\ee
These weights can be seen as associated to the faces, and not to the vertices of $\Delta^*$. Therefore, there is no non-trivial dynamics associated with the vertices.

Instead, if we use the weak imposition of the simplicity constraints, following the EPRL prescription, to compute and constrain the partition function of the $\SO(4)$ $BF$ theory, we obtain the state sum
\ba
\mathcal{Z}_{(\mathcal{C},\Phi)}
&=&\sum_{\{j\}\rightarrow\{f\}}\prod_{f\in\Delta^*}(2j_f+1)^2\vcenter{\hbox{\includegraphics[scale=0.3]{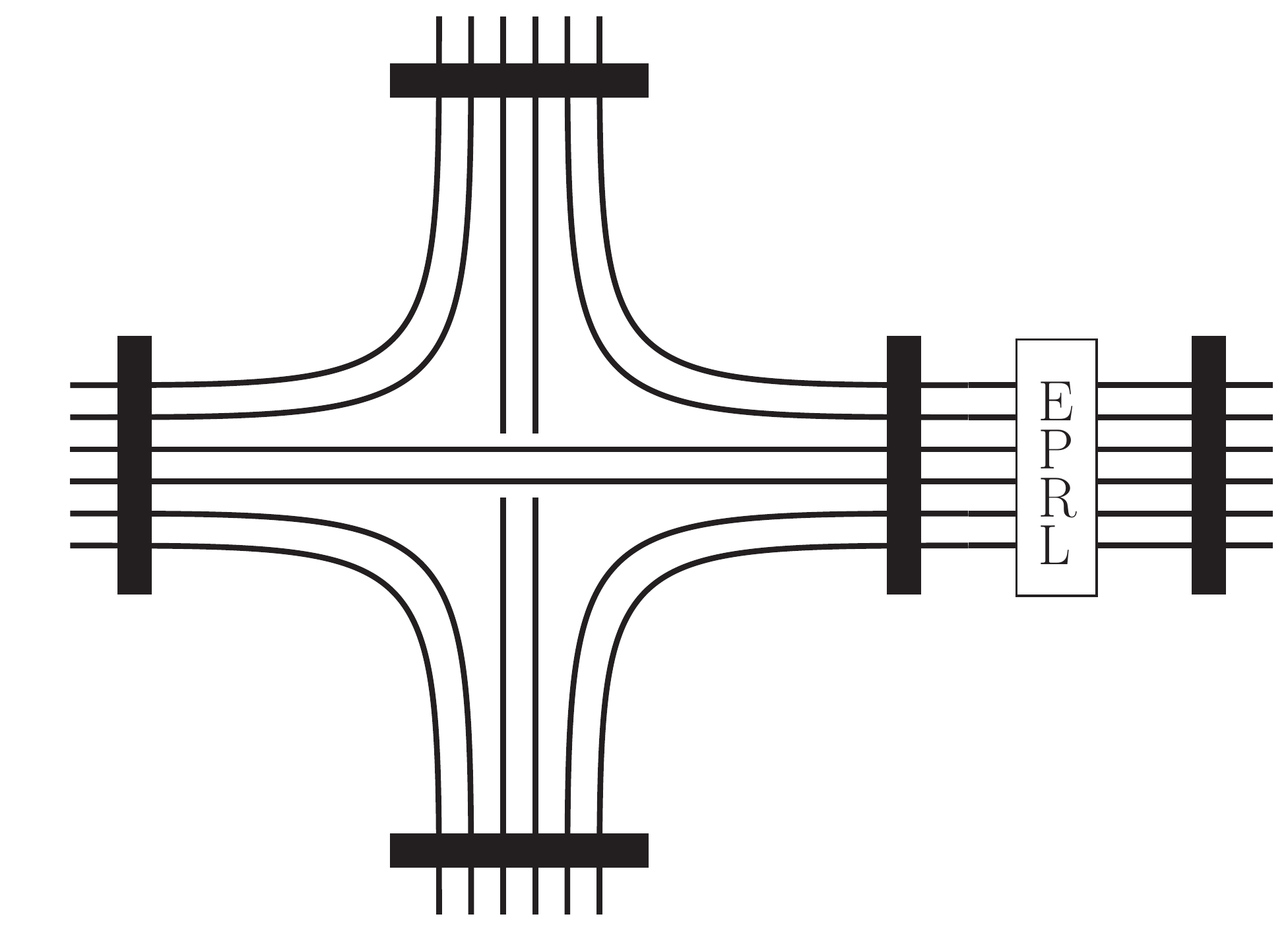}}}\nonumber\\
&=&\sum_{\{j\}\rightarrow\{f\}}\prod_{f\in\Delta^*}(2j_f+1)^2\vcenter{\hbox{\includegraphics[scale=0.3]{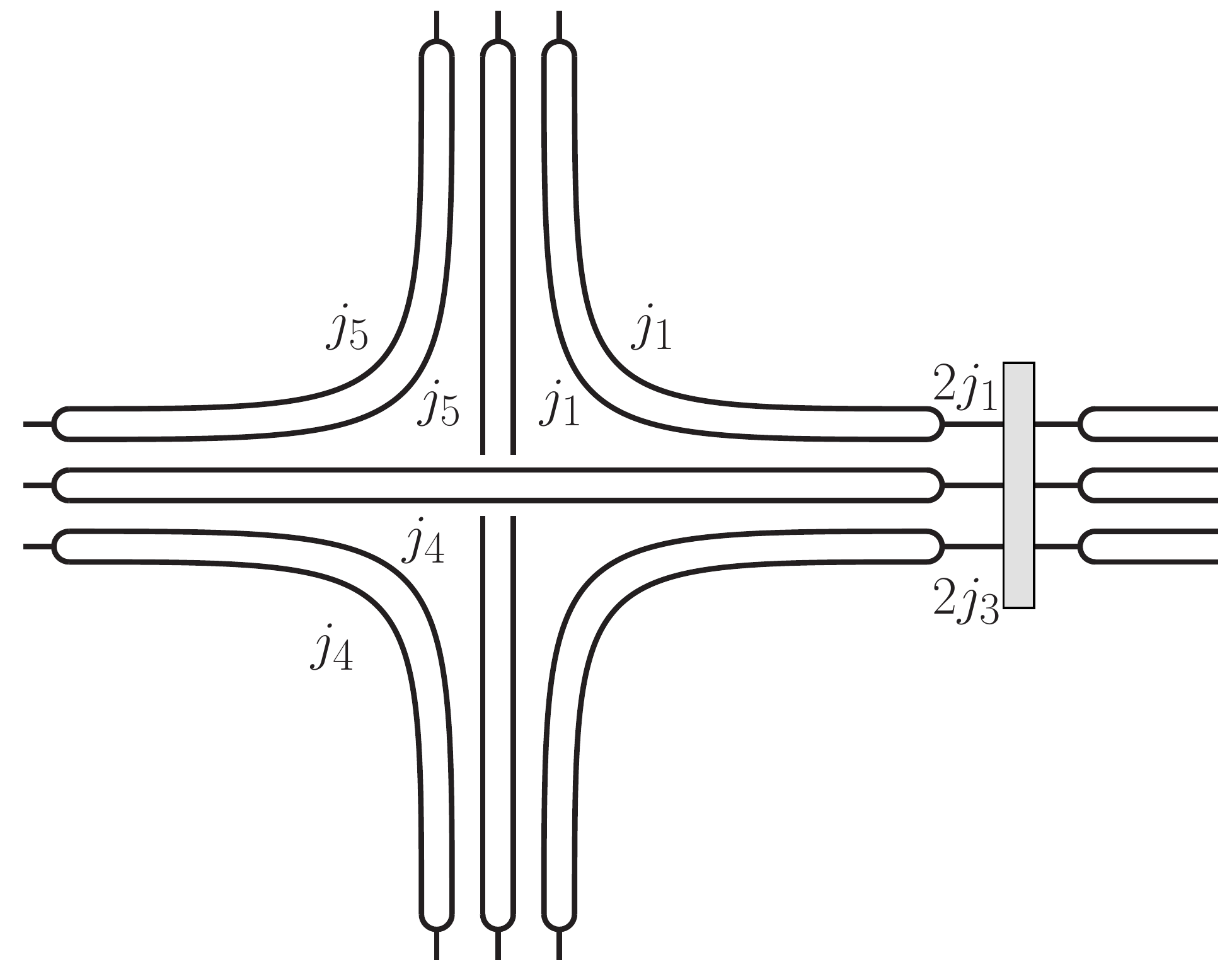}}}\nonumber
\ea
\newpage
\ba
\textcolor{white}{\mathcal{Z}_{\mathcal{C},\Phi}}
&=&\sum_{\{j\}\rightarrow\{f\}}\prod_{f\in\Delta^*}(2j_f+1)^2\vcenter{\hbox{\includegraphics[scale=0.3]{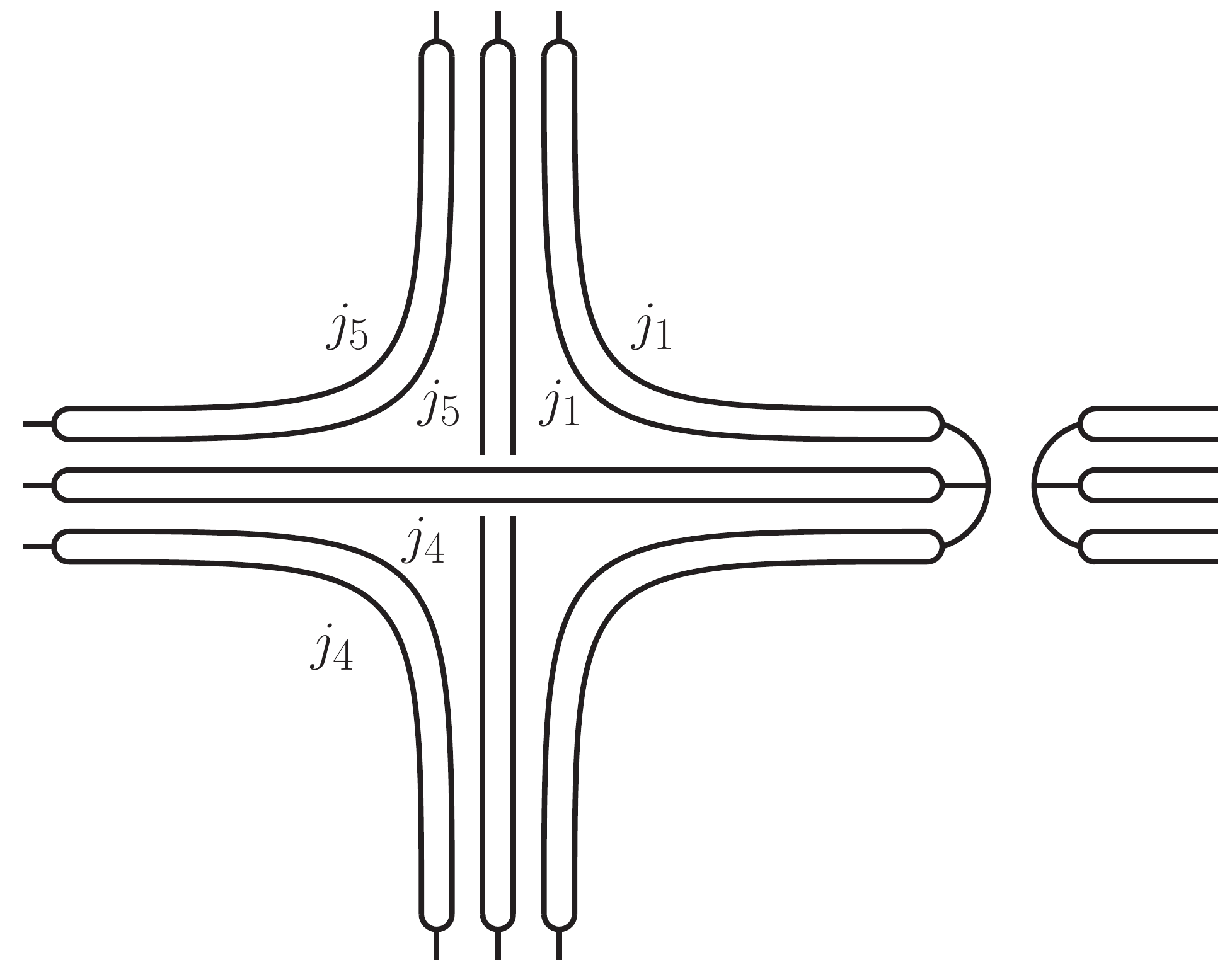}}}\nonumber\\
&=&\sum_{\{j\}\rightarrow\{f\}}\prod_{f\in\Delta^*}(2j_f+1)^2\prod_{v\in\Delta^*}
\f{1}{d_{j_1}\dots d_{j_6}}
\left\{
\begin{array}{ccc}
2j_1&2j_2&2j_3\\
2j_4&2j_5&2j_6
\end{array}\right\}.\nonumber
%\vcenter{\hbox{\includegraphics[scale=0.3]{VertexNEW4.pdf}}}.
\ea
The vertex amplitude is now given by a $6j$ symbol, which is what we expected.

\subsection{Discussion}

\noindent Let us make a few comments about the result obtained above. The first one concerns the weights associated with the faces. By taking into account the secondary second class constraints, we did not recover the correct weights for the faces. As we already said, this comes from the fact that we should have used the correct non-trivial path integral measure. It would be interesting to check wether taking into account this measure leads to the correct face amplitude.

The second remark concerns the fact that the $6j$ symbol associated with the vertices involves the representations $2j_f$ and not the representations $j_f$. This might very well be a consequence of the fact that the state sum model that we have constructed corresponds classically to two copies of the Einstein-Hilbert action, according to (\ref{einstein-hilbert}).

Finally, we would like to discuss the relevance of the EPRL prescription compared to the BC one. Both prescriptions are constraints on the intertwiner degrees of freedom labeling the edges of $\Delta^*$, while the secondary second class constraints that we have imposed act on the faces. As we have seen, even with a proper implementation of the secondary second class constraints, the BC prescription is too strong and ``kills'' all the dynamics. It is therefore clear that the simplicity constraints $\mathcal{C}$ should be imposed weakly, which is an argument in favor of the so-called new models.

\section{Note on the inclusion of the Barbero-Immirzi parameter}
\label{sec3}

\noindent We would like to end this work with a note about the Barbero-Immirzi parameter $\gamma$. In four-dimensional canonical loop quantum gravity, the Barbero-Immirzi parameter has been introduced in order to allow for the construction of the real (in the sense of the field of real numbers) $\su(2)$-valued Ashtekar-Barbero connection \cite{barbero,immirzi}. It is well known that it has no effects at the classical level (see \cite{fermions1,fermions2,fermions3,fermions4,fermions5} for a discussion about its controversial fate in the presence of fermions), since it drops out of the equations of motion by means of the Bianchi identities. However, in the quantum theory, it appears explicitly  in the spectra of the geometric operators \cite{ALvolume,rovelli-smolin} such as area and volume, and in the formula for the entropy of black holes \cite{rovelli-black hole,ABK,meissner,agullo,ENP}. It has been argued that the presence of this quantization ambiguity at the quantum level might be related to a ``wrong'' choice of the connection parametrizing the phase space at the classical level \cite{alexandrov-vassilevich}.

In the context of four-dimensional spin foam models, the inclusion of the Barbero-Immirzi parameter has been made possible with the introduction of the EPRL and FK models. This has opened the possibility of relating the boundary states of spin foam models with the kinematical states of loop quantum gravity. However, several points still lack a clear explanation. In particular, the structure of the models seem to depend on the particular values taken by $\gamma$. For $\gamma<1$, the EPRL and FK models do coincide \cite{EPRL}, and the space of boundary states can be reduced to the space of $\SU(2)$ spin networks. For $\gamma>1$ however, the two models are different, and even if the boundary states of the EPRL model coincide with the states of canonical loop quantum gravity, the space of boundary states of the FK model seems to be bigger.

Since we have given in this work a proposal for a rigorous implementation of the spin foam quantization procedure in the three-dimensional context, we believe that it should be possible and interesting to investigate the issues arising in the presence of the Barbero-Immirzi parameter. In usual three-dimensional $\SU(2)$ Riemannian gravity, such an extension of the classical theory is not possible, since there is only a unique (invariant non-degenerate) bilinear form on $\su(2)$. For this reason, previous attempts to introduce a Barbero-Immirzi-like parameter were relying on the Chern-Simons formulation of $BF$ theory with $\Lambda\neq0$, in which case there is another bilinear form available \cite{bonzom-livine}. However, this does not mimic what happens in the four-dimensional theory, and renders difficult the comparison between the two situations.

Since the three-dimensional Plebanski theory (\ref{3d plebanski}) is formulated for the gauge group $\SO(4)$, it can be generalized in a natural way to the case $\gamma\neq0$. Indeed, since there are two traces, denoted by $\Tr(\cdot\,,\cdot)$ and $\Tr(\star\,\cdot\,,\cdot)$, available on $\so(4)$, it is possible to define two different pairings of elements of $\so(4)$, and to construct the modified action
\be\nonumber
S_\text{BF$\gamma$}[B,\omega]
=\int_\mathcal{M}\de^3x\,\eps^{\mu\nu\rho}\left(\Tr(B_\mu,F_{\nu\rho})+\f{1}{\gamma}\Tr(\star B_\mu,F_{\nu\rho})\right),
\ee
where $\gamma\in\mathbb{R}-\{0\}$. To see that $\gamma$ is the analogue of the four-dimensional Barbero-Immirzi parameter, and in particular that it disappears on-shell, let us split the $B$ field into its self-dual and anti self-dual components, and write the action as
\be\nonumber
S_\text{BF$\gamma$}[B,\omega]=\left(1+\f{1}{\gamma}\right)\p S_\text{BF$\gamma$}+\left(1-\f{1}{\gamma}\right)\m S_\text{BF$\gamma$},
\ee
where
\be\nonumber
\PM S_\text{BF$\gamma$}[\PM B,\PM\omega]\equiv\int_\mathcal{M}\de^3x\,\eps^{\mu\nu\rho}\,\Tr(\PM B_\mu,\PM F_{\nu\rho}).
\ee
Following what we have done in section \ref{sec1}, it is possible to solve the torsion-free condition (if $\PM B_\mu$ is invertible), and to plug back the solution for the connection in the action $S_\text{BF$\gamma$}$ in order to obtain the second order Einstein-Hilbert action
\be\nonumber
\epsilon^+\left(1+\f{1}{\gamma}\right)S_\text{EH}[\p g_{\mu\nu}]+\epsilon^-\left(1-\f{1}{\gamma}\right)S_\text{EH}[\m g_{\mu\nu}].
\ee
If we impose that $B^{IJ}_\mu$ be simple (i.e. if we choose the gravitational sector of solutions to the simplicity constraints), we know from (\ref{selfdual BG}) and (\ref{urbantke metrics}) that we have $\epsilon^+=\epsilon^-$ and $\p g_{\mu\nu}=\m g_{\mu\nu}$, and therefore it is clear the the Barbero-Immirzi parameter $\gamma$ disappears from the classical theory.

This argument shows that it is possible to naturally extend the three-dimensional Plebanski action that we have introduced in order to have a non-vanishing Barbero-Immirzi parameter which behaves classically exactly as in the four-dimensional theory. The Holst-Plebanski action
\be\nonumber
S_\text{Pl$\gamma$}[B,\omega,\phi]
=\f{1}{2}\int_\mathcal{M}\de^3x\left[\eps^{\mu\nu\rho}\left(\Tr(B_\mu,F_{\nu\rho})+\f{1}{\gamma}\Tr(\star B_\mu,F_{\nu\rho})\right)
+\phi^{\mu\nu}\,\Tr(B_\mu,\star B_\nu)\right]
\ee
can now be taken as a starting point for the spin foam quantization. It is clear that $\gamma$ plays no role at the classical level, and we can then study its fate in the quantum theory. It has been claimed in the four-dimensional context that a canonical quantization based on the non-commutative spacetime Lorentz-connection leads to a quantum theory independent of $\gamma$ \cite{alexandrov4}. It is therefore natural to ask if $\gamma$ can disappear as well in the spin foam quantization, and, if not, to investigate why it is not the case. We think that this question deserves a detailed treatment, and keep it for future work. We want to stress that this is the first time that a Barbero-Immirzi parameter is introduced in three-dimensional gravity.

\section*{Discussion and conclusion}

\noindent In this work, we have introduced a three-dimensional Plebanski action for the gauge group $\SO(4)$, which is based on the notion of simplicity for one-forms. This simplicity condition on the $B$ field is imposed via quadratic simplicity constraints, which admit, just like in the four-dimensional case, a gravitational, a topological, and a degenerate sector of solutions. We have shown at the classical level that the action reduces to the Einstein-Hilbert action in the gravitational sector, while in the topological sector the theory is trivial. The canonical analysis reveals that the theory is indeed equivalent to three-dimensional gravity. Interestingly, this model has a Hamiltonian structure similar to that of the four-dimensional Plebanski theory, and in particular features secondary second class constraints arising from the stabilization of the primary simplicity constraints. 

This three-dimensional Plebanski action allows to test the imposition of the simplicity constraints in the spin foam quantization of gravity. We show that neither the BC nor the EPRL prescriptions lead to the expected vertex amplitude, which is that of the Ponzano-Regge model. However, the two schemes differ drastically in that only the EPRL model can give the proper amplitude when it is supplemented with the imposition of the secondary second class constraints. This suggests that the weak imposition of the simplicity constraints is an essential ingredient in the spin foam quantization, but that it is not sufficient in this three-dimensional model. We have given a concrete realization of the idea that imposing the secondary second class constraints is a missing element of the spin foam approach \cite{alexandrov1}.

A limitation of the present model is that, because of its three-dimensional nature, it has trivial off-diagonal simplicity constraints, and therefore the BC and EPRL prescriptions lead to the same vertex amplitude (up to a normalization factor). This is a key difference with the four-dimensional case, where the off-diagonal simplicity constraints are non-trivial, and might very well encode the information about the secondary second class constraints. In order to investigate this possibility, it would be interesting to extend the present three-dimensional model to an arbitrary non-simplicial cellular decomposition.

Interestingly, the three-dimensional $\SO(4)$ Plebanski action allows for the introduction of a Barbero-Immirzi parameter. This could be a good starting point to understand and test issues related to the role of the Barbero-Immirzi parameter in the canonical and spin foam quantizations \cite{geiller-noui-gamma}.

Finally, let us conclude with a remark on the BC model. In the present work, we have followed only one very specific way to derive the BC model, which is the most straightforward one. There are however in the literature more rigorous and transparent approaches which lead to variants of the usual four-dimensional BC model, and which allow to overcome some of the criticism which were originally raised against the model and to clarify its geometrical meaning. Details about these constructions can be found in the work of Bonzom and Livine \cite{bonzom-livine} and Baratin and Oriti \cite{baratin-oriti}. In this latter in particular, the authors use a non-commutative metric representation of group field theory to define a model of four-dimensional constrained $BF$ theory, whose amplitudes reproduce a variant of that of the BC model. In this approach, it is shown that any spin foam model for constrained $BF$ theory can be written as a simplicial path integral featuring an explicit non-trivial measure over the holonomies and the bivectors. The measure over the holonomies is naturally interpreted as imposing secondary constraints arising from the consistent imposition in each tetrahedron frame of the (primary) linear simplicity constraints. It would therefore be interesting to investigate wether these secondary constraints agree with those coming from the canonical analysis, and to test the proposal of \cite{baratin-oriti,baratin-oriti2} on the present toy model to see if it leads to the Ponzano-Regge amplitudes.

\section*{Acknowledgments}

\noindent It is a pleasure to thank Alejandro Perez for useful comments, discussions, and enthusiasm. We also would like to thank Sergei Alexandrov and Simone Speziale for reading an earlier version of this work, and pointing out interesting remarks. Finally, we thank Aristide Baratin for discussions about group field theory.

\appendix
\section{Notations}
\label{notations}

\noindent In this work, notations are such that $\mu,\nu,\dots$ refer to spacetime indices, $a,b,\dots$ to spatial indices, $I,J,\dots$ to $\so(4)$ indices, and $i,j,\dots$ to $\su(2)$ indices. We assume that the three-dimensional spacetime manifold $\mathcal{M}$ is topologically $\Sigma\times\mathbb{R}$, where $\Sigma$ is a two-dimensional manifold without boundaries. We use the notation $\lambda^I$ for vectors in $\mathbb{R}^4$ with components $(\lambda^0,\lambda^i)$, the wedge product between elements $u,v\in\mathbb{R}^3$ (zero-forms or one-forms) to denote the operation $(u\wedge v)^i=\eps^{i}_{~jk}u^jv^k$, and the dot product for $u\cdot v=u^iv_i$. We often denote the vectors $\lambda^i\in\mathbb{R}^3$ simply by $\lambda$.

\section{The Lie algebra $\boldsymbol{\so(4)}$}
\label{so4}

\noindent $\so(4)$ is the real Lie algebra of the isometry group $\SO(4)$ of the quadratic form $\delta=\text{diag}(1,1,1,1)$. We use capital Latin letters for internal vector indices and define the antisymmetric tensor $\eps^{IJKL}$ such that $\eps^{0123}=1$ and $\eps_{IJKL}=\eps^{IJKL}$. The indices are lowered and raised with the metric $\delta$. The action of the Hodge dual operator is defined by
\be\nonumber
\star J^{IJ}=\f{1}{2}\eps^{IJ}_{~~KL}J^{KL},
\ee
and it satisfies $\star^2=\text{id}$.

One of the basis of $\so(4)$, composed of rotation generators $L_i$ and boost generators $K_i$, with $i\in\lbrace1,2,3\rbrace$, has the following commutation relations:
\be\nonumber
[L_i,L_j]=\eps_{ij}^{~~k}L_k,\qquad[K_i,K_j]=\eps_{ij}^{~~k}L_k,\qquad[K_i,L_j]=\eps_{ij}^{~~k}K_k,
\ee
where $\eps_{ijk}\equiv\eps^0_{~ijk}$. Starting from this basis of $\so(4)$, it is convenient to define a new basis $\PM J_i$ as
\be\nonumber
\PM J_i=\f{1}{2}(L_i\pm K_i),
\ee
whose generators realize two commuting copies of $\su(2)$, i.e. satisfy
\be\nonumber
[\PM J_i,\PM J_j]=\eps_{ij}^{~~k}\PM J_k,\qquad[\p J_i,\m J_j]=0.
\ee
For any element $\xi\in\so(4)$, we have the decomposition $\xi=\p\xi+\m\xi$ into self-dual and anti self-dual components, where $\PM\xi=\PM\xi^i\PM J_i$. The action of the Hodge dual operator on the (anti) self-dual components if given by
\be\nonumber
\star\PM\xi=\pm\PM\xi.
\ee
Finally, the vector representation of $\so(4)$ is given by
\be\nonumber
\PM J^{IJ}_i=\f{1}{2}\left(-\eps^{0iIJ}\pm\delta^{iI}\delta^{0J}\mp\delta^{0I}\delta^{iJ}\right).
\ee

\section{Sign of $\boldsymbol{\det(\PM B)}$ in the topological and gravitational sectors}
\label{appdet}

\noindent In this appendix, we compute the sign $\epsilon^\pm$ of $\det(\PM B)$ in the gravitational and topological sectors.

\subsection{Gravitational sector}

\noindent In the gravitational sector, we have $B^{IJ}_\mu=\eps^{IJ}_{~~KL}\chi^Ke^L_\mu$, and the boost and rotational components of $B^{IJ}_\mu$ are given respectively by
\be\nonumber
K^i_\mu\equiv B^{0i}_\mu=(\chi\wedge e_\mu)^i,
\qquad
L^i_\mu\equiv\f{1}{2}\eps^i_{~jk}B^{jk}_\mu=\chi^0e^i_\mu-\chi^ie^0_\mu.
\ee
Therefore, we have the relation
\be\nonumber
K^i_\mu=\chi_0^{-1}(\chi\wedge L_\mu)^i,
\ee
and the self-dual and anti self-dual components (\ref{selfdual BG}) of the $B$ field can be written as
\be\label{BdecompositionG}
\PM B^i_\mu=\mp(\chi\wedge e_\mu)^i+(\chi^ie^0_\mu-\chi^0e^i_\mu)=\mp K^i_\mu-L^i_\mu=(-\mathbb{I}\mp\chi_0^{-1}\underline{\chi})L^i_\mu.
\ee
Here we have introduced the three-dimensional matrix
\be\nonumber
\underline{\chi}=
 \begin{pmatrix}
0 & -\chi_3 & \chi_2\\
\chi_3 & 0 & -\chi_1\\
-\chi_2 & \chi_1 & 0
 \end{pmatrix}
\ee
associated to $\chi$ such that $\underline{\chi}\alpha^i=(\chi\wedge\alpha)^i$ for any $\alpha\in\mathbb{R}^3$, and $\mathbb{I}$ denotes the three-dimensional unit matrix. With this notation, we can compute from (\ref{BdecompositionG}) the determinant
\be\nonumber
\det(\PM B)=\det(-\mathbb{I}\mp\chi_0^{-1}\underline{\chi})\det(L^i_\mu)=-\big(1+\chi_0^{-2}(\chi_1^2+\chi_2^2+\chi_3^2)\big)\det(L^i_\mu).
\ee
Therefore, we see that in the gravitational sector we have $\epsilon^+=\epsilon^-$.

\subsection{Topological sector}

\noindent In the topological sector, we have $B^{IJ}_\mu=\star\eps^{IJ}_{~~KL}\chi^Ke^L_\mu=\chi^Ie^J_\mu-\chi^Je^I_\mu$, and the boost and rotational components of $B^{IJ}_\mu$ are given respectively by
\be\nonumber
K^i_\mu=\chi^0e^i_\mu-\chi^ie^0_\mu,
\qquad
L^i_\mu=(\chi\wedge e_\mu)^i.
\ee
Therefore, we have the relation
\be\nonumber
L^i_\mu=\chi_0^{-1}(\chi\wedge K_\mu)^i,
\ee
and the self-dual and anti self-dual components (\ref{selfdual BT}) of the $B$ field can be written as
\be\nonumber
\PM B^i_\mu=-(\chi\wedge e_\mu)^i\pm(\chi^ie^0_\mu-\chi^0e^i_\mu)=\mp K^i_\mu-L^i_\mu=(\mp\mathbb{I}-\chi_0^{-1}\underline{\chi})K^i_\mu.
\ee
From this formula, we can compute the determinant
\be\nonumber
\det(\PM B)=\det(\mp\mathbb{I}-\chi_0^{-1}\underline{\chi})\det(K^i_\mu)=\mp\big(1+\chi_0^{-2}(\chi_1^2+\chi_2^2+\chi_3^2)\big)\det(K^i_\mu).
\ee
Therefore, we see that in the topological sector we have $\epsilon^+=-\epsilon^-$.

\section{Algebra of constraints}
\label{constraints algebra}

\noindent The algebra of constraints is
\ba
%&&\lb T(u),T(v)\rb=-T([u,v]),\nonumber\\
%&&\lb T(u),\mathcal{K}_0\rb=0,\nonumber\\
%&&\lb T(u),\mathcal{K}_a\rb=\Tr(\pi_0,[u,B_a])\approx0,\nonumber\\
%&&\lb T(u),\Psi_\mu\rb\approx0,\nonumber\\
%&&\lb T(u),\widetilde{\mathcal{K}}_0\rb=0,\nonumber\\
%&&\lb T(u),\widetilde{\mathcal{K}}_a\rb=\Tr(\star\pi_0,[u,B_a])\approx0,\nonumber\\
%&&\lb T(u),\Phi_{ab}\rb=\Tr([\star u,B_0],\mathcal{D}_aB_b)+\Tr([\star u,B_0],\mathcal{D}_aB_b),\nonumber\\
%&&\lb T(u),\mathcal{C}_{00}\rb=0,\nonumber\\
%&&\lb T(u),\mathcal{C}_{0a}\rb=\Tr(\star B_0,[u,B_a]),\nonumber\\
%&&\lb T(u),\mathcal{C}_{ab}\rb=0,\nonumber\\
&&\lb\mathcal{K}_0,\mathcal{K}_0\rb=0,\nonumber\\
&&\lb\mathcal{K}_0,\mathcal{K}_a\rb=\Tr(\pi_0,B_a)\approx0,\nonumber\\
&&\lb\mathcal{K}_a,\mathcal{K}_b\rb=0,\nonumber\\\cr
&&\lb\mathcal{K}_0,\Psi_0\rb=-\Psi_0\approx0,\nonumber\\
&&\lb\mathcal{K}_0,\Psi_a\rb=0,\nonumber\\
&&\lb\mathcal{K}_a,\Psi_0\rb=-\Psi_a+2\Tr(\pi_0,\mathcal{D}_aB_0)\approx0,\nonumber\\
&&\lb\mathcal{K}_a,\Psi_b\rb=2\Tr(\pi_0,\mathcal{D}_aB_b)\approx0,\nonumber\\\cr
&&\lb\mathcal{K}_0,\widetilde{\mathcal{K}}_0\rb=0,\nonumber\\
&&\lb\mathcal{K}_0,\widetilde{\mathcal{K}}_a\rb=\Tr(\pi_0,\star B_a)\approx0,\nonumber\\
&&\lb\mathcal{K}_a,\widetilde{\mathcal{K}}_0\rb=-\Tr(B_a,\star\,\pi_0)\approx0,\nonumber\\
&&\lb\mathcal{K}_a,\widetilde{\mathcal{K}}_b\rb=0,\nonumber\\\cr
&&\lb\mathcal{K}_0,\Phi_{ab}\rb=-\Tr(\mathcal{D}_aB_0,\star B_b)-\Tr(\mathcal{D}_bB_0,\star B_a)=-\Phi_{ab}\approx0,\nonumber\\
&&\lb\mathcal{K}_c,\Phi_{ab}\rb=-\Tr(\mathcal{D}_aB_c,\star B_b)-\Tr(\mathcal{D}_bB_c,\star B_a)
								+\eps_{ac}\Tr(\pi_0,[B_0,\star B_b])+\eps_{bc}\Tr(\pi_0,[B_0,\star B_a])\approx0,\nonumber\\\cr
&&\lb\mathcal{K}_\mu,\mathcal{C}_{00}\rb=-2\Tr(B_\mu,\star B_0)=-2\mathcal{C}_{\mu0}\approx0,\nonumber\\
&&\lb\mathcal{K}_\mu,\mathcal{C}_{0a}\rb=-\Tr(B_\mu,\star B_a)=-\mathcal{C}_{\mu a}\approx0,\nonumber\\
&&\lb\mathcal{K}_\mu,\mathcal{C}_{ab}\rb=0,\nonumber\\\cr
&&\lb\Psi_0,\Psi_0\rb=0,\nonumber\\
&&\lb\Psi_0,\Psi_a\rb=-2\eps^{cd}\,\Tr(F_{cd},\mathcal{D}_aB_0),\nonumber\\
&&\lb\Psi_a,\Psi_b\rb=2\eps^{cd}\,\Tr(F_{cd},\mathcal{D}_aB_b)-2\eps^{cd}\,\Tr(F_{cd},\mathcal{D}_bB_a),\nonumber\\\cr
&&\lb\Psi_0,\widetilde{\mathcal{K}}_0\rb=\eps^{ab}\,\Tr(F_{ab},\star B_0),\nonumber\\
&&\lb\Psi_0,\widetilde{\mathcal{K}}_c\rb=\eps^{ab}\,\Tr(F_{ab},\star B_c)-2\Tr(\mathcal{D}_c\pi_0,\star B_0)\approx\eps^{ab}\,\Tr(F_{ab},\star B_c),\nonumber\\
&&\lb\Psi_a,\widetilde{\mathcal{K}}_\mu\rb\approx0,\nonumber\\\cr
&&\lb\Psi_0,\Phi_{ab}\rb=-2\Tr(\mathcal{D}_aB_0,\star\mathcal{D}_bB_0)-2\Tr(\mathcal{D}_bB_0,\star\mathcal{D}_aB_0),\nonumber\\
&&\lb\Psi_c,\Phi_{ab}\rb=-2\Tr(\mathcal{D}_aB_c,\star\mathcal{D}_bB_0)-2\Tr(\mathcal{D}_bB_c,\star\mathcal{D}_aB_0)
						-\eps^{de}\Tr(F_{de},\eps_{ca}[B_0,\star B_b]+\eps_{cb}[B_0,\star B_a]),\nonumber\\\cr
&&\lb\Psi_\mu,\mathcal{C}_{00}\rb=0,\nonumber\\
&&\lb\Psi_\mu,\mathcal{C}_{0a}\rb=2\Tr(\star B_\mu,\mathcal{D}_aB_0),\nonumber\\
&&\lb\Psi_\mu,\mathcal{C}_{ab}\rb=2\Tr(\star B_\mu,\mathcal{D}_aB_b)+2\Tr(\star B_\mu,\mathcal{D}_bB_a).\nonumber
\ea

\end{document}